\newcommand{\MSbar}{\overline{\rm MS}}
\newcommand{\deltaMSbar}{\delta_{\overline{\rm MS}}}
\newcommand{\DRbar}{\overline{\rm DR}}
\newcommand{\DRbarprime}{\overline{\rm DR}'}
\newcommand{\lnbar}{{\overline{\rm ln}}}
\newcommand\beq{\begin{eqnarray}}
\newcommand\eeq{\end{eqnarray}}
\newcommand{\propA}{{\rm A}}   
\newcommand{\propB}{{\rm B}}   
\newcommand{\propS}{{\rm S}}
\newcommand{\propU}{{\rm U}}
\newcommand{\propM}{{\rm M}}
\newcommand{\propV}{{\rm V}}
\newcommand{\propW}{{\rm W}}
\newcommand{\propX}{{\rm X}}
\newcommand{\propY}{{\rm Y}}
\newcommand{\propZ}{{\rm Z}}
\newcommand{\squark}{\tilde q}

\newcommand{\quark}{q}

\newcommand{\gluino}{\tilde g}
\newcommand{\Fbar}{{\overline F}}

\documentclass[twocolumn,
amsmath,
prd,nofootinbib,floatfix
]{revtex4}

\allowdisplaybreaks
\usepackage{axodraw}
\usepackage{graphicx}
\usepackage{bm}

\begin{document}
\renewcommand{\theequation}{\arabic{section}.\arabic{equation}}

\title{Two-loop scalar self-energies and pole masses in a general\\
renormalizable theory with massless gauge bosons}

\author{Stephen P. Martin}
\affiliation{
Physics Department, Northern Illinois University, DeKalb IL 60115 USA\\
{\rm and}
Fermi National Accelerator Laboratory, PO Box 500, Batavia IL 60510}

\phantom{.}

\begin{abstract} I present the two-loop self-energy functions for scalar
bosons in a general renormalizable theory, within the approximation that
vector bosons are treated as massless or equivalently that gauge
symmetries are unbroken.  This enables the computation of the two-loop
physical pole masses of scalar particles in that approximation. The
calculations are done simultaneously in the mass-independent $\MSbar$,
$\DRbar$, and $\DRbarprime$ renormalization schemes, and with arbitrary
covariant gauge fixing.  As an example, I present the two-loop SUSYQCD
corrections to squark masses, which can increase the known one-loop
results by of order one percent.  More generally, it is now
straightforward to implement all two-loop sfermion pole mass computations
in supersymmetry using the results given here, neglecting only the
electroweak vector boson masses compared to the superpartner masses in the
two-loop parts.

\end{abstract}


\maketitle

\tableofcontents

\section{Introduction}\label{sec:introduction}
\setcounter{equation}{0}
\setcounter{footnote}{1}

Low-energy supersymmetry provides a way of understanding the small ratio
of the electroweak breaking scale in the Standard Model to other very high
energy scales. This requires the existence of complex scalar superpartners
for each of the known quarks and leptons, with masses not far above 1 TeV.
These squarks and sleptons should be accessible to the Fermilab Tevatron
$p\overline p$ collider or the CERN Large Hadron Collider, and their
masses can be measured with refined precision at a future linear $e^+e^-$
collider; see for example \cite{LHCILC} and references therein. The match
between these measurements and particular models of supersymmetry breaking
will then require calculations at the two-loop level of precision, at
least.  An important part of this is the calculation of self-energy
functions, which can in turn be used to calculate the physical masses of
the new particles. 

In general, the mass given by the position of the complex pole in the
propagator is a gauge-invariant and renormalization scale-invariant
quantity \cite{Tarrach:1980up}-\cite{Gambino:1999ai}. The pole mass does
suffer from ambiguities \cite{poleambiguities} due to infrared physics
associated with the QCD confinement scale, but these are probably not
large enough to cause a practical problem for strongly-interacting
superpartners.  The pole mass should be closely related in a calculable
way to the kinematic observable mass reported by experiments
\cite{massdefs}. In recent years, many important higher-order calculations
of self-energy functions and pole masses in the Standard Model have been
performed, including two-loop \cite{Gray:1990yh}-\cite{Jegerlehner:2003py}
and three-loop \cite{Chetyrkin:1999qi}-\cite{Melnikov:2000qh}
contributions for quarks, and two-loop results for electroweak vector
bosons \cite{Chang:1981qq}-\cite{Jegerlehner:2001fb}, as well as two-loop
results for top and bottom quarks in supersymmetry \cite{quarkpoleSUSY}. 

In a previous paper \cite{Martin:2003it}, I provided partial results for
the two-loop self-energy functions for scalars in a general renormalizable
theory, using the approximation that no more than one vector boson
propagator is included. That is a useful approximation for the Higgs
scalar boson(s), for which the most important contributions at the
two-loop level involve the strong interactions and/or Yukawa couplings. In
this paper, I will extend the previous result by including the
contributions for any number of vector boson lines, within the
approximation that the gauge symmetry is unbroken so that the vector
bosons are massless. Because of the experimental exclusions of light
sfermions already achieved by the CERN LEP $e^+e^-$ collider
\cite{LEPSUSY} and the Fermilab Tevatron $p\overline p$ collider
\cite{squarksDzero,squarksCDF}, this will likely give a very good
approximation for the squark and slepton pole masses.  (Here, the effects
of non-zero $W$ and $Z$ boson masses can be included as usual in the
one-loop part \cite{Pierce:1996zz}, and neglected in the two-loop part.)


\section{Notations and Setup\label{sec:setup}}
\setcounter{equation}{0}
\setcounter{footnote}{1}

Let us write the tree-level squared-mass eigenstate fields of the theory
as\footnote{Since a complex scalar can be written as two real scalars, 
and a Dirac fermion as two Weyl fermions, this entails no loss of generality.} 
a set of real scalars $R_i$, two-component
Weyl fermions $\psi_I$, and vector bosons $V^\mu_A$. Scalar field indices
are $i,j,k,\ldots$, fermion flavor indices are $I,J,K,\ldots$, and
$A,B,C,\ldots$ run over the adjoint representation of the gauge group,
while $\mu,\nu,\ldots$ are space-time vector indices.
Repeated indices of all types are summed over unless otherwise noted.
The kinetic part of the Lagrangian is taken to be:
\beq
{\cal L}_{\rm kin} &=&
-\frac{1}{2} \partial_\mu R_i \partial^\mu R_i 
- \frac{1}{2} m^2_i R_i^2
\nonumber \\ 
&&
-i\psi^{\dagger I} \overline \sigma^\mu \partial_\mu \psi_I - 
\frac{1}{2} (M^{IJ} \psi_I \psi_J + {\rm c.c.})
\nonumber \\ 
&& 
-\frac{1}{2} (\partial_\mu V_\nu^A - \partial_\nu V_\mu^A)
\partial^\mu V^{\nu}_A .
\eeq
The metric tensor has signature ($-$$+$$+$$+$).
The non-gauge interactions of these fields are given by:
\beq
{\cal L}_{\rm int} &=& 
-\frac{1}{6} \lambda^{ijk} R_i R_j R_k
-\frac{1}{24} \lambda^{ijkm} R_i R_j R_k R_m
\nonumber \\ 
&&
-\frac{1}{2} (y^{JKi} \psi_J \psi_K R_i + {\rm c.c.}).
\eeq
where $\lambda^{ijk}$ and $\lambda^{ijkm}$ are real couplings and the Yukawa
couplings $y^{JKi}$ are symmetric complex matrices on the indices $J,K$, 
for each $i$. Raising or lowering of fermion indices implies 
complex conjugation, so
\beq
M_{IJ} \equiv (M^{IJ})^*,\qquad y_{JKi} \equiv (y^{JKi})^* .
\eeq
The heights of real scalar and vector indices have no significance, and
are chosen for convenience.
The scalar squared masses $m_i^2$ and the fermion squared
masses $M_{IK} M^{KJ} = m^2_I \delta_I^J$ are taken to have been 
diagonalized (by an appropriate rotation of the fields if necessary).
However, the fermion mass matrix $M^{IJ}$ is not necessarily
diagonal, but must have non-zero entries only when $I$ and $J$ label 
two-component fermions with the same squared mass and in conjugate
representations of the gauge group.

In order to completely specify the pertinent features of the
gauge interactions of the theory, 
let $T^A$ be the Hermitian generator matrices
of the gauge group for a (possibly reducible) representation $R$.
They are labeled by an adjoint representation index $A$ corresponding
to the vector bosons of the theory, $V_A^\mu$.
They satisfy $[T^A, T^B] = i f^{ABC} T^C$, where $f^{ABC}$ are
the totally antisymmetric structure constants of the gauge group.
Then results below are written in terms of the invariants:
\beq
{(T^A T^A)_i}^j &=& C(i) \delta_i^j,\\
{\rm Tr}[T^A T^B] &=& I(R) \delta^{AB},\\
f^{ACD} f^{BCD} &=& C(G) \delta^{AB},
\eeq
which define the quadratic Casimir invariant for the representation 
carrying the index $i$, the total Dynkin index summed over the
representation $R$, and the Casimir invariant of the adjoint representation
of the group,
respectively. When the gauge group contains several simple or $U(1)$
factors labeled $a,b,c,\ldots$ with distinct gauge couplings $g_a$,
the corresponding invariants are written $C_a(i)$, $I_a(R)$, and
$C_a(G)$. The normalization is such that for $SU(N)$, $C(G) = N$ and each 
fundamental 
representation
has $C(i) = (N^2-1)/2N$ and contributes $1/2$ to $I(R)$.
For a $U(1)$ gauge group, $C(G) = 0$ and a representation with charge $q$ 
has $C(i) = q^2$
and contributes $q^2$ to $I(R)$. 
The results given below will be presented
in terms of these group theory invariants for the representations carried
by the scalar and fermion degrees of freedom.

The computations in this paper are performed in a general gauge with a
vector boson propagator obtained by covariant
gauge fixing in the usual way:
\beq
-i\delta_{AB} [\eta^{\mu\nu}/k^2 + (\xi -1) k^\mu k^\nu/(k^2)^2] .
\label{eq:vectorprop}
\eeq
Here $\xi=0$ for Landau gauge and $\xi=1$ for Feynman gauge and
$\xi=3$ for the Fried-Yennie gauge \cite{FriedYennie}, for a
vector boson carrying 4-momentum $k^\mu$.
Infrared divergences are dealt with by first computing with a finite
vector boson mass, and later taking the massless vector limit. 
All contributions involving gauge boson loops
implicitly include the corresponding contributions of ghost loops.

For each Feynman diagram, the integrations over internal momenta are
regulated by continuing to $d = 4 - 2 \epsilon$ dimensions, according to
\begin{eqnarray}
\int d^4 k \rightarrow (2 \pi \mu)^{2 \epsilon} \int d^d k .
\end{eqnarray}
In the dimensional regularization scheme, the vector bosons also have $d$
components, while in the dimensional reduction scheme they have $d$
ordinary components and $2\epsilon$ additional components known as epsilon
scalars. 
For the present case of massless vector bosons, this means that 
the 4-dimensional metric in the vector propagator of 
eq.~(\ref{eq:vectorprop}) is replaced by
\beq
\eta^{\mu\nu}/k^2 &\rightarrow&
g^{\mu\nu}/k^2 + \hat g^{\mu\nu}/(k^2 + m_\epsilon^2) ,
\label{eq:decomposemetric}
\eeq
where $g^{\mu\nu}$ is projected onto a formal $d$--dimensional
subspace, and $\hat g^{\mu\nu}$ onto the complementary 
$2\epsilon$--dimensional subspace.
Counterterms for
the one-loop sub-divergences and the remaining two-loop divergences are
added, according to the rules of minimal subtraction, to give finite 
results, which then depend on the renormalization scale $Q$ given by
\begin{eqnarray}
Q^2 = 4 \pi e^{-\gamma} \mu^2 .
\end{eqnarray}
Logarithms of dimensionful quantities are always written in terms of
\beq
\lnbar X \equiv \ln (X/Q^2) 
.
\eeq
The resulting renormalization schemes are known as $\MSbar$ \cite{MSbar}
and $\DRbar$ \cite{DRbar}, respectively, for the cases 
in which $\hat g^{\mu\nu}$ is not and is included. 

The epsilon-scalar squared mass parameter $m_\epsilon^2$ appearing in the
$\DRbar$ scheme is unphysical.
One could set $m_\epsilon^2$ equal to zero at any fixed renormalization
scale, but then it will be non-zero at other renormalization scales,
since it has a non-homogeneous beta function \cite{Jack:1994kd}. 
Furthermore, under renormalization group evolution it will feed into the 
ordinary scalar squared masses in the $\DRbar$ scheme. 
Fortunately, a
redefinition (given in \cite{DRbarprime} at one-loop order, and at two-loop
order in \cite{effpot}) of the ordinary scalar squared masses 
completely removes the dependence on the unphysical
epsilon scalar squared mass $m_\epsilon^2$ from the renormalization group
equations and the equations relating tree-level parameters to physical
observables. The resulting $\DRbarprime$ scheme \cite{DRbarprime} 
is therefore appropriate for softly-broken supersymmetric theories
such as the Minimal Supersymmetric Standard Model (MSSM).
In this paper, calculations will be presented simultaneously in all three 
schemes, using the following two devices. First,
\begin{eqnarray}
\deltaMSbar \equiv \begin{cases}
1 & \text{for}\quad\MSbar \\
0 & \text{for}\quad\DRbar,\> \DRbarprime .
\end{cases}
\label{eq:deltaMSbar}
\end{eqnarray}
Second, terms that involve the unphysical parameter
$m_\epsilon^2$ should be construed below to apply
only to the $\DRbar$ scheme, not the $\DRbarprime$ or $\MSbar$ schemes.

A main objective of this paper is to compute the two-loop scalar
self-energy
\begin{eqnarray}
\Pi_{ij}(s) =
\frac{1}{16\pi^2} \Pi_{ij}^{(1)} +
\frac{1}{(16\pi^2)^2} \Pi_{ij}^{(2)} + \ldots,
\end{eqnarray}  
a (complex, in general) symmetric matrix, as a function of
\begin{eqnarray}
s = -p^2,
\label{eq:defines}
\end{eqnarray}
where $p^\mu$ is the external momentum.
Note that $s$ is taken to be real
with an infinitesimal positive imaginary part to resolve the branch cuts.
The self-energy function $\Pi_{ij}$ is 
calculated as the sum of connected, one-particle 
irreducible, two-point Feynman diagrams.
It is gauge-dependent, but can be used to obtain a
gauge-invariant physical
squared mass, defined as the position of the complex pole, with   
non-positive imaginary part, in the propagator obtained from the
perturbative Taylor expansion of the self-energy function. For scalar
particles with tree-level renormalized (running) squared masses $m_k^2$, 
the two-loop pole squared masses
\begin{eqnarray}
s_{k} = M_k^2 - i \Gamma_k M_k
\end{eqnarray}  
are obtained as the solutions to
\begin{eqnarray}
\mbox{Det}\left [ (m_i^2- s_{k})
\delta_{ij} +  \Pi_{ij} (s_k) \right ] = 0.
\label{eq:genpole}
\end{eqnarray}
A gauge-invariant and renormalization scale invariant solution 
at two-loop order is obtained by first expanding the self-energy
in a Taylor series in $s$ about the tree-level squared mass, with the 
result for the complex pole mass:
\beq
s_k &=& m_k^2 + \frac{1}{16\pi^2} \Pi^{(1)}_{kk} +
 \frac{1}{(16\pi^2)^2} \Bigl [  \Pi^{(2)}_{kk}
+  \Pi^{(1)}_{kk}  \Pi^{(1)\prime}_{kk}
\phantom{xxx}
\nonumber \\ 
&&
+  \sum_{j\not=k} (\Pi^{(1)}_{kj})^2/(m_k^2 - m_j^2) \Bigr ] ,
\label{eq:poleexp}
\eeq
where the prime indicates differentiation with respect to $s$, and all
self-energy functions on the right-hand side are evaluated with $s
\rightarrow m_k^2$. This assumes that, as is the case for example for
sfermions in the MSSM, the scalars that mix with each other are not
degenerate, so that the last term is a well-defined part of a perturbative
expansion. If (nearly) degenerate scalars do mix, then the appropriate
version of (nearly) degenerate perturbation theory should be used instead.

One can also obtain a solution iteratively, by first taking 
$s_k = m_k^2$ as
the argument of the self-energy eq.~(\ref{eq:genpole}), and then taking
the resulting value for $s_k$ and substituting it in as the argument of
the self-energy function, repeating the process until sufficient numerical
convergence is obtained. In this case, since $s_k$ has a negative
imaginary part on the physical sheet and the argument $s$ of the
self-energy is taken to be real with a positive imaginary part, the
self-energy for complex $s$ can be defined in terms of its Taylor series
expansion about a point on the real $s$ axis.
However, for a theory with massless gauge bosons, the terms of a given 
loop order in the expansion of the self-energy have branch cuts.
For example, at one loop order,
\beq
\Pi^{(1)}_{ij}(s) = (\xi-3) g_a^2 C_a(i) \delta_{ij} m_i^2 P(s/m_i^2)
+ \ldots
\eeq
where
\beq
P(x) = (x - 1/x) \ln (1-x - i\varepsilon) ,
\eeq
and the ellipses refers to terms without branch cuts. This yields a
result that is not perturbative in the gauge coupling, unless one
takes the Fried-Yennie gauge-fixing condition $\xi=3$. So, although
the pole mass is formally gauge invariant, this iterative procedure
has quite poor convergence unless $\xi=3$. At least in the examples given
below in section \ref{sec:examples}, I find that the iterative procedure
in Fried-Yennie gauge gives good agreement with eq.~(\ref{eq:poleexp}), 
the difference being formally of three-loop order in any case, and the
implementation of eq.~(\ref{eq:poleexp}) is simpler 
and computationally faster. The checks of gauge invariance and 
renormalization scale invariance for particular examples
are also obtained most straightforwardly
from eq.~(\ref{eq:poleexp}).

The results below will be written in terms of two-loop integral basis 
functions, following the notation given in \cite{evaluation,TSIL}.
The one-loop and two-loop integral functions are reduced using 
Tarasov's algorithm \cite{Tarasov:1997kx,Mertig:1998vk}
to a set of
basis integrals $A(x)$, $B(x,y)$, $I(x,y,z)$, $S(x,y,z)$, $T(x,y,z)$, 
$U(x,y,z,u)$,
and $M(x,y,z,u,v)$, corresponding to the Feynman diagram
topologies shown in fig.~1. 
Here $x,y,z,u,v$ are squared mass arguments, and 
the arguments $s$ and $Q^2$ are not shown explicitly, because 
they are 
the same for all functions in a given equation. 
The name of a particle stands for its squared mass when appearing as
an argument of a loop-integral function. A prime on an argument of
one of these functions indicates a derivative with respect to that
argument, so that $T(x,y,z) = -S(x',y,z) = -\partial S(x,y,z)/\partial x$.
Also, $I(x,y,z) = S(x,y,z)|_{s=0}$. It is often useful to
define the functions $V(x,y,z,u) 
= -U(x,y',z,u)$,
and ${\overline T} (0,y,z) = \lim_{x \rightarrow 0} 
[T(x,y,z) + B(y,z) \lnbar x]$. 
These and $B(x,y')$ can be reduced to combinations of other
basis functions, but they
arise quite often in applications in such a way that explicit
reduction would needlessly complicate the expressions.
The basis integrals contain counterterms that render them
ultraviolet finite. The precise definitions, and the
calculation of these functions and a publicly available 
computer code for that purpose, are described in 
\cite{evaluation,TSIL}.

The one-loop and two-loop Feynman diagrams in the approximation used in
this paper are shown in fig.~\ref{fig:alldiagrams}, labeled according to
a convention described in detail in ref.~\cite{Martin:2003it}. The results
of their evaluations are described in the next section.

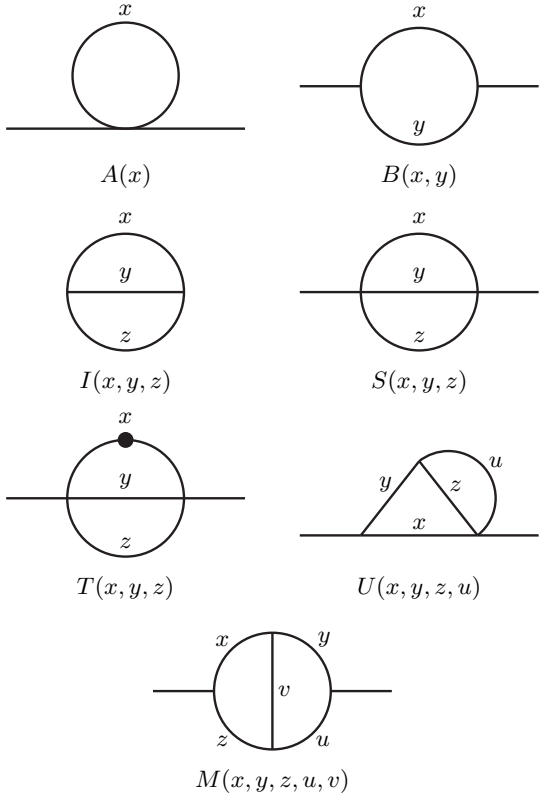
\begin{figure}[t]  

\begin{picture}(108,76)(-54,-40)
\SetWidth{0.9}
\CArc(0,4)(20,0,360)
\Text(0,29)[]{$x$}
\Text(0,-34)[]{$A(x)$}
\Line(-45,-16)(45,-16)
\end{picture}
\begin{picture}(108,76)(-54,-40)
\SetWidth{0.9}
\Line(-45,0)(-22,0)
\Line(45,0)(22,0)
\CArc(0,0)(22,0,360)
\Text(0,28)[]{$x$}
\Text(0,-16)[]{$y$}
\Text(0,-34)[]{$B(x,y)$}
\end{picture}
\begin{picture}(108,77)(-54,-40)
\SetWidth{0.9}
\CArc(0,0)(22,0,360)
\Line(-22,0)(22,0)
\Text(0,28)[]{$x$}
\Text(0,6.5)[]{$y$}
\Text(0,-17)[]{$z$}
\Text(0,-34)[]{$I(x,y,z)$}
\end{picture}
\begin{picture}(108,77)(-54,-40)
\SetWidth{0.9}
\Line(-45,0)(-22,0)
\Line(45,0)(22,0)
\CArc(0,0)(22,0,360)
\Line(-22,0)(22,0)
\Text(0,28)[]{$x$}
\Text(0,6.5)[]{$y$}
\Text(0,-17)[]{$z$}
\Text(0,-34)[]{$S(x,y,z)$}
\end{picture}
\begin{picture}(108,77)(-54,-40)
\SetWidth{0.9}
\Line(-45,0)(-22,0)   
\Line(45,0)(22,0)  
\CArc(0,0)(22,0,360)
\Line(-22,0)(22,0)   
\Vertex(0,22){3.0}
\Text(0,30.2)[]{$x$}  
\Text(0,6.5)[]{$y$}
\Text(0,-17)[]{$z$}
\Text(0,-34)[]{$T(x,y,z)$}
\end{picture}
\begin{picture}(108,72)(-54,-40)
\SetWidth{0.9}
\Line(-45,-14)(-22,-14)
\Line(45,-14)(22,-14)
\Line(-22,-14)(0,14)
\Line(22,-14)(0,14)
\CArc(11,0)(17.8045,-51.843,128.157)
\Line(-22,-14)(22,-14)
\Text(0,-9.67)[]{$x$}
\Text(-12.8,5.3)[]{$y$}
\Text(13.8,4.8)[]{$z$}
\Text(29,13.6)[]{$u$}
\Text(0,-34)[]{$U(x,y,z,u)$}
\end{picture}
\begin{picture}(108,72)(-54,-40)
\SetWidth{0.9}
\Line(-45,0)(-22,0)
\Line(45,0)(22,0)
\CArc(0,0)(22,0,360) 
\Line(0,-22)(0,22) 
\Text(-18.9,18.9)[]{$x$}
\Text(19.7,19.7)[]{$y$}   
\Text(-19,-19)[]{$z$}
\Text(19,-19)[]{$u$}
\Text(4.5,0)[]{$v$}   
\Text(0,-34)[]{$M(x,y,z,u,v)$}
\end{picture}
\caption{\label{fig:topologies} 
Feynman diagram topologies for the one- and two-loop 
self-energy basis integrals used in this paper. The dot on the
$T$ diagram means that $T(x,y,z) = -\partial S(x,y,z)/\partial x$.
The precise definitions of the integral functions
and methods for their 
evaluation are described in \cite{evaluation,TSIL}.}
\end{figure}

\begin{figure*}[p] 
\SetScale{0.66}

\begin{picture}(98,64)(-49,-35)
\SetWidth{0.9}
\DashLine(-45,-22)(0,-22){4}
\DashLine(45,-22)(0,-22){4}
\DashCArc(0,0)(22,-90,270){4}
\Text(0,-32)[]{$\propA_{S}$}
\end{picture}
\hspace{-0.3in}
\begin{picture}(98,64)(-49,-35)
\SetWidth{0.9}
\DashLine(-45,0)(-22,0){4}
\DashLine(45,0)(22,0){4}
\DashCArc(0,0)(22,0,180){4}
\DashCArc(0,0)(22,180,360){4}
\Text(0,-32)[]{$\propB_{SS}$}
\end{picture}
\hspace{-0.3in}
\begin{picture}(98,64)(-49,-35)
\SetWidth{0.9}
\DashLine(-45,0)(-22,0){4}
\DashLine(45,0)(22,0){4}
\CArc(0,0)(22,0,180)
\CArc(0,0)(22,180,360)
\Text(0,-32)[]{$\propB_{FF}$}
\end{picture}
\hspace{-0.3in}
\begin{picture}(98,64)(-49,-35)
\SetWidth{0.9}
\DashLine(-45,0)(-22,0){4}
\DashLine(45,0)(22,0){4}
\PhotonArc(0,0)(22,0,180){2}{8.5}
\DashCArc(0,0)(22,180,360){4}
\Text(0,-32)[]{$\propB_{SV}$}
\end{picture}
\hspace{-0.3in}
\begin{picture}(98,64)(-49,-35)
\SetWidth{0.9}
\DashLine(-45,-22)(0,-22){4}
\DashLine(45,-22)(0,-22){4}
\PhotonArc(0,0)(22,-90,270){-2}{14.5}
\Text(0,-32)[]{$\propA_{V} \>\> (*)$}
\end{picture}
\hspace{-0.3in}
\begin{picture}(98,64)(-49,-35)
\SetWidth{0.9}
\DashLine(-45,0)(-22,0){4}
\DashLine(45,0)(22,0){4}
\PhotonArc(0,0)(22,0,180){2}{8.5}
\PhotonArc(0,0)(22,180,360){2}{8.5}
\Text(0,-32)[]{$\propB_{VV}$}
\end{picture}

\vspace{-8pt}

\begin{picture}(98,82)(-49,-39)
\SetWidth{0.9}
\DashLine(-45,-22)(-20,-22){4}
\DashLine(45,-22)(20,-22){4}
\DashLine(-20,-22)(20,-22){4}
\DashLine(-20,-22)(0,7){4}
\DashLine(20,-22)(0,7){4}
\DashCArc(0,22)(15,-90,270){4}
\Text(0,-32)[]{$\propY_{SSSS}$}
\end{picture}
\hspace{-0.3in}
\begin{picture}(98,82)(-49,-39)
\SetWidth{0.9}
\DashLine(-45,-22)(0,-22){4}
\DashLine(45,-22)(0,-22){4}
\DashCArc(0,12)(22,0,180){4}
\DashLine(-22,12)(22,12){4}
\DashLine(-22,12)(22,12){4}
\DashLine(0,-22)(-22,12){4}
\DashLine(0,-22)(22,12){4}
\Text(0,-32)[]{$\propW_{SSSS}$}
\end{picture}
\hspace{-0.3in}
\begin{picture}(98,82)(-49,-39)
\SetWidth{0.9}
\DashLine(-45,-22)(0,-22){4}
\DashLine(45,-22)(0,-22){4}
\DashCArc(0,-7)(15,-90,90){4}
\DashCArc(0,-7)(15,90,270){4}
\DashCArc(0,23)(15,-90,270){4}
\Text(0,-32)[]{$\propX_{SSS}$}
\end{picture}
\hspace{-0.3in}
\begin{picture}(98,82)(-49,-39)
\SetWidth{0.9}
\DashLine(-45,0)(-22,0){4}
\DashLine(45,0)(22,0){4}
\DashCArc(0,0)(22,0,180){4}
\DashCArc(0,0)(22,180,360){4}
\DashLine(0,-22)(0,22){4}
\Text(0,-32)[]{$\propM_{SSSSS}$}
\end{picture}
\hspace{-0.3in}
\begin{picture}(98,82)(-49,-39)
\SetWidth{0.9}
\DashLine(-45,-22)(-22,-22){4}
\DashLine(45,-22)(22,-22){4}
\DashLine(-22,-22)(22,-22){4}
\DashCArc(0,12)(22,0,180){4}
\DashLine(-22,12)(22,12){4}
\DashLine(-22,-22)(-22,12){4}
\DashLine(22,-22)(22,12){4}
\Text(0,-32)[]{$\propV_{SSSSS}$}
\end{picture}
\hspace{-0.3in}
\begin{picture}(98,82)(-49,-39)
\SetWidth{0.9}
\DashLine(-48,0)(-30,0){4}
\DashLine(48,0)(30,0){4}
\DashCArc(-15,0)(15,0,180){4}
\DashCArc(-15,0)(15,180,360){4}
\DashCArc(15,0)(15,0,180){4}
\DashCArc(15,0)(15,180,360){4}
\Text(0,-32)[]{$\propZ_{SSSS}$}
\end{picture}

\vspace{-8pt}

\begin{picture}(98,82)(-49,-39)
\SetWidth{0.9}
\DashLine(-45,-14)(-22,-14){4}
\DashLine(45,-14)(22,-14){4}
\DashLine(-22,-14)(0,14){4}
\DashLine(22,-14)(0,14){4}
\DashCArc(11,0)(17.8045,-51.843,128.157){4}
\DashLine(-22,-14)(22,-14){4}
\Text(0,-32)[]{$\propU_{SSSS}$}
\end{picture}
\hspace{-0.3in}
\begin{picture}(98,82)(-49,-39)
\SetWidth{0.9}
\DashLine(-45,0)(-22,0){4}
\DashLine(45,0)(22,0){4}
\DashCArc(0,0)(22,0,180){4}
\DashCArc(0,0)(22,180,360){4}
\DashLine(-22,0)(22,0){4}
\Text(0,-32)[]{$\propS_{SSS}$}
\end{picture}
\hspace{-0.3in}
\begin{picture}(98,82)(-49,-39)
\SetWidth{0.9}
\DashLine(-45,0)(-22,0){4}
\DashLine(45,0)(22,0){4}
\CArc(0,0)(22,0,180)
\CArc(0,0)(22,180,360)
\DashLine(0,-22)(0,22){4}
\Text(0,-32)[]{$\propM_{FFFFS}$}
\end{picture}
\hspace{-0.3in}
\begin{picture}(98,82)(-49,-39)
\SetWidth{0.9}
\DashLine(-45,-22)(-22,-22){4}
\DashLine(45,-22)(22,-22){4}
\Line(-22,-22)(22,-22)
\DashCArc(0,12)(22,0,180){4}
\Line(-22,12)(22,12)
\Line(-22,-22)(-22,12)
\Line(22,-22)(22,12)
\Text(0,-32)[]{$\propV_{FFFFS}$}
\end{picture}
\hspace{-0.3in}
\begin{picture}(98,82)(-49,-39)
\SetWidth{0.9}
\DashLine(-45,0)(-22,0){4}
\DashLine(45,0)(22,0){4}
\CArc(0,0)(22,-90,90)
\DashCArc(0,0)(22,90,180){4}
\DashCArc(0,0)(22,180,270){4}
\Line(0,-22)(0,22)
\Text(0,-32)[]{$\propM_{SFSFF}$}
\end{picture}
\hspace{-0.3in}
\begin{picture}(98,82)(-49,-39)
\SetWidth{0.9}
\DashLine(-45,-22)(-22,-22){4}
\DashLine(45,-22)(22,-22){4}
\Line(-22,12)(22,12)
\CArc(0,12)(22,0,180)
\DashLine(-22,-22)(22,-22){4}
\DashLine(-22,-22)(-22,12){4}
\DashLine(22,-22)(22,12){4}
\Text(0,-32)[]{$\propV_{SSSFF}$}
\end{picture}

\vspace{-8pt}

\begin{picture}(98,82)(-49,-39)
\SetWidth{0.9}
\DashLine(-45,-22)(0,-22){4}
\DashLine(45,-22)(0,-22){4}
\CArc(0,12)(22,0,180)
\Line(-22,12)(22,12)
\DashLine(0,-22)(-22,12){4}
\DashLine(0,-22)(22,12){4}
\Text(0,-32)[]{$\propW_{SSFF}$}
\end{picture}
\hspace{-0.3in}
\begin{picture}(98,82)(-49,-39)
\SetWidth{0.9}
\DashLine(-45,-22)(-20,-22){4}
\DashLine(45,-22)(20,-22){4}
\Photon(-20,-22)(20,-22){2}{5.5}
\DashLine(-20,-22)(0,7){4}
\DashLine(20,-22)(0,7){4}
\DashCArc(0,22)(15,-90,270){4}
\Text(0,-32)[]{$\propY_{VSSS}$}
\end{picture}
\hspace{-0.3in}
\begin{picture}(98,82)(-49,-39)
\SetWidth{0.9}
\DashLine(-45,-22)(0,-22){4}
\DashLine(45,-22)(0,-22){4}
\PhotonArc(0,12)(22,0,180){2}{7.5}
\DashLine(-22,12)(22,12){4}
\DashLine(-22,12)(22,12){4}
\DashLine(0,-22)(-22,12){4}
\DashLine(0,-22)(22,12){4}
\Text(0,-32)[]{$\propW_{SSSV}$}
\end{picture}
\hspace{-0.3in}
\begin{picture}(98,82)(-49,-39)
\SetWidth{0.9}
\DashLine(-45,-22)(0,-22){4}
\DashLine(45,-22)(0,-22){4}
\DashCArc(0,-7)(15,-90,90){4}
\DashCArc(0,-7)(15,90,270){4}
\PhotonArc(0,23)(15,-90,270){-2}{10.5}
\Text(0,-32)[]{$\propX_{SSV} \>\> (*)$}
\end{picture}
\hspace{-0.3in}
\begin{picture}(98,82)(-49,-39)
\SetWidth{0.9}
\DashLine(-45,0)(-22,0){4}
\DashLine(45,0)(22,0){4}
\DashCArc(0,0)(22,0,180){4}
\DashCArc(0,0)(22,180,360){4}
\Photon(0,-22)(0,22){2}{5}
\Text(0,-32)[]{$\propM_{SSSSV}$}
\end{picture}
\hspace{-0.3in}
\begin{picture}(98,82)(-49,-39)
\SetWidth{0.9}
\DashLine(-45,-22)(-22,-22){4}
\DashLine(45,-22)(22,-22){4}
\DashLine(-22,-22)(22,-22){4}
\PhotonArc(0,12)(22,0,180){2}{7.5}
\DashLine(-22,12)(22,12){4}
\DashLine(-22,-22)(-22,12){4}
\DashLine(22,-22)(22,12){4}
\Text(0,-32)[]{$\propV_{SSSSV}$}
\end{picture}

\vspace{-9pt}

\begin{picture}(98,82)(-49,-39)
\SetWidth{0.9}
\DashLine(-45,0)(-22,0){4}
\DashLine(45,0)(22,0){4}
\DashCArc(0,0)(22,-90,90){4}
\PhotonArc(0,0)(22,90,180){2}{4.5}
\DashCArc(0,0)(22,180,270){4}
\DashLine(0,-22)(0,22){4}
\Text(0,-32)[]{$\propM_{VSSSS}$}
\end{picture}
\hspace{-0.3in}
\begin{picture}(98,82)(-49,-39)
\SetWidth{0.9}
\DashLine(-45,-22)(-22,-22){4}
\DashLine(45,-22)(22,-22){4}
\DashLine(-22,12)(22,12){4}
\DashCArc(0,12)(22,0,180){4}
\Photon(-22,-22)(22,-22){-2}{5.5}
\DashLine(-22,-22)(-22,12){4}
\DashLine(22,-22)(22,12){4}
\Text(0,-32)[]{$\propV_{VSSSS}$}
\end{picture}
\hspace{-0.3in}
\begin{picture}(98,82)(-49,-39)
\SetWidth{0.9}
\DashLine(-45,-22)(-20,-22){4}
\DashLine(45,-22)(20,-22){4}
\DashLine(-20,-22)(20,-22){4}
\DashLine(-20,-22)(0,7){4}
\DashLine(20,-22)(0,7){4}
\PhotonArc(0,22)(15,-90,270){-2}{10.5}
\Text(0,-32)[]{$\propY_{SSSV}\>\> (*)$}
\end{picture}
\hspace{-0.3in}
\begin{picture}(98,82)(-49,-39)
\SetWidth{0.9}
\DashLine(-45,0)(-22,0){4}
\DashLine(45,0)(22,0){4}
\CArc(0,0)(22,0,180)
\CArc(0,0)(22,180,360)
\Photon(0,-22)(0,22){2}{5}
\Text(0,-32)[]{$\propM_{FFFFV}$}
\end{picture}
\hspace{-0.3in}
\begin{picture}(98,82)(-49,-39)
\SetWidth{0.9}
\DashLine(-45,-22)(-22,-22){4}
\DashLine(45,-22)(22,-22){4}
\Line(-22,-22)(22,-22)
\PhotonArc(0,12)(22,0,180){2}{7.5}
\Line(-22,12)(22,12)
\Line(-22,-22)(-22,12)
\Line(22,-22)(22,12)
\Text(0,-32)[]{$\propV_{FFFFV}$}
\end{picture}
\hspace{-0.3in}
\begin{picture}(98,82)(-49,-39)
\SetWidth{0.9}
\DashLine(-45,0)(-22,0){4}
\DashLine(45,0)(22,0){4}
\CArc(0,0)(22,-90,90)
\PhotonArc(0,0)(22,90,180){2}{4.5}
\DashCArc(0,0)(22,180,270){4}
\Line(0,-22)(0,22)
\Text(0,-32)[]{$\propM_{VFSFF}$}
\end{picture}

\vspace{-8pt}

\begin{picture}(98,82)(-49,-39)
\SetWidth{0.9}
\DashLine(-45,-22)(-22,-22){4}
\DashLine(45,-22)(22,-22){4}
\Line(-22,12)(22,12)
\CArc(0,12)(22,0,180)
\Photon(-22,-22)(22,-22){-2}{5.5}
\DashLine(-22,-22)(-22,12){4}
\DashLine(22,-22)(22,12){4}
\Text(0,-32)[]{$\propV_{VSSFF}$}
\end{picture}
\hspace{-0.3in}
\begin{picture}(98,82)(-49,-39)
\SetWidth{0.9}
\DashLine(-45,-22)(-22,-22){4}
\DashLine(45,-22)(22,-22){4}
\Photon(-22,-22)(22,-22){-2.2}{5.5}
\PhotonArc(0,12)(22,0,180){2.2}{7.5}
\DashLine(-22,12)(22,12){4}
\DashLine(-22,-22)(-22,12){4}
\DashLine(22,-22)(22,12){4}
\Text(0,-32)[]{$\propV_{VSSSV}$}
\end{picture}
\hspace{-0.3in}
\begin{picture}(98,82)(-49,-39)
\SetWidth{0.9}
\DashLine(-45,-22)(-20,-22){4}
\DashLine(45,-22)(20,-22){4}
\Photon(-20,-22)(20,-22){-2}{5.5}
\DashLine(-20,-22)(0,7){4}
\DashLine(20,-22)(0,7){4}
\PhotonArc(0,22)(15,-90,270){-2}{10.5}
\Text(0,-32)[]{$\propY_{VSSV} \>\> (*)$}
\end{picture}
\hspace{-0.3in}
\begin{picture}(98,82)(-49,-39)
\SetWidth{0.9}
\DashLine(-48,0)(-30,0){4}
\DashLine(48,0)(30,0){4}
\PhotonArc(-15,0)(15,0,180){-1.7}{6}
\DashCArc(-15,0)(15,180,360){4}
\PhotonArc(15,0)(15,0,180){1.7}{6}
\DashCArc(15,0)(15,180,360){4}
\Text(0,-32)[]{$\propZ_{VSVS}$}
\end{picture}
\hspace{-0.3in}
\begin{picture}(98,82)(-49,-39)
\SetWidth{0.9}
\DashLine(-45,-14)(-22,-14){4}
\DashLine(45,-14)(22,-14){4}
\DashLine(-22,-14)(0,14){4}
\DashLine(22,-14)(0,14){4}
\PhotonArc(11,0)(17.8045,-51.843,128.157){2}{7.5}
\Photon(-22,-14)(22,-14){-2}{5.5}
\Text(0,-32)[]{$\propU_{VSSV}$}
\end{picture}
\hspace{-0.3in}
\begin{picture}(98,82)(-49,-39)
\SetWidth{0.9}
\DashLine(-45,0)(-22,0){4}
\DashLine(45,0)(22,0){4}
\PhotonArc(0,0)(22,0,180){2}{8.5}
\DashCArc(0,0)(22,180,360){4}
\Photon(-22,0)(22,0){2}{5.5}
\Text(0,-32)[]{$\propS_{SVV}$}
\end{picture}

\vspace{-8pt}

\begin{picture}(98,82)(-49,-39)
\SetWidth{0.9}
\DashLine(-45,0)(-22,0){4}
\DashLine(45,0)(22,0){4}
\DashCArc(0,0)(22,0,90){4}
\PhotonArc(0,0)(22,90,180){2.3}{4.5}
\DashCArc(0,0)(22,180,270){4}
\PhotonArc(0,0)(22,270,360){2.3}{4.5}
\DashLine(0,-22)(0,22){4}
\Text(0,-32)[]{$\propM_{VSSVS}$}
\end{picture}
\hspace{-0.3in}
\begin{picture}(98,82)(-49,-39)
\SetWidth{0.9}
\DashLine(-45,-22)(-22,-22){4}
\DashLine(45,-22)(22,-22){4}
\DashLine(-22,-22)(22,-22){4}
\Photon(-22,-22)(-22,12){2.5}{4.5}
\Photon(22,-22)(22,12){-2.5}{4.5}
\Photon(-22,12)(22,12){-2.5}{5.5}
\PhotonArc(0,12)(22,0,180){2.2}{7.5}
\Text(0,-32)[]{$\propV_{SVVVV}$}
\end{picture}
\hspace{-0.3in}
\begin{picture}(98,82)(-49,-39)
\SetWidth{0.9}
\DashLine(-45,-22)(0,-22){4}
\DashLine(45,-22)(0,-22){4}
\PhotonArc(0,12)(22,0,180){2.2}{7.5}
\Photon(-22,12)(22,12){2.2}{5.5}
\Photon(0,-22)(-22,12){2}{4.5}
\Photon(0,-22)(22,12){-2}{4.5}
\Text(0,-32)[]{$\propW_{VVVV} \>\> (*)$}
\end{picture}
\hspace{-0.3in}
\begin{picture}(98,82)(-49,-39)
\SetWidth{0.9}
\DashLine(-45,-22)(-20,-22){4}
\DashLine(45,-22)(20,-22){4}
\DashLine(-20,-22)(20,-22){4}
\Photon(-20,-22)(0,7){2}{4.5}
\Photon(20,-22)(0,7){-2}{4.5}
\PhotonArc(0,22)(15,-90,270){-2}{10.5}
\Text(0,-32)[]{$\propY_{SVVV} \>\> (*)$}
\end{picture}
\hspace{-0.3in}
\begin{picture}(98,82)(-49,-39)
\SetWidth{0.9}
\DashLine(-45,-22)(0,-22){4}
\DashLine(45,-22)(0,-22){4}
\PhotonArc(0,-7)(15,-90,90){-1.8}{5.5}
\PhotonArc(0,-7)(15,90,270){-1.8}{5.5}
\PhotonArc(0,23)(15,-90,270){-2}{10.5}
\Text(0,-32)[]{$\propX_{VVV}\>\> (*)$}
\end{picture}
\hspace{-0.3in}
\begin{picture}(98,82)(-49,-39)
\SetWidth{0.9}
\DashLine(-45,0)(-22,0){4}
\DashLine(45,0)(22,0){4}
\PhotonArc(0,0)(22,0,90){2.5}{4.5}
\PhotonArc(0,0)(22,90,180){2.5}{4.5}
\DashCArc(0,0)(22,180,270){4}
\DashCArc(0,0)(22,270,360){4}
\Photon(0,-22)(0,22){2.6}{6}
\Text(0,-32)[]{$\propM_{VVSSV}$}
\end{picture}

\vspace{-9pt}

\begin{picture}(98,82)(-49,-39)
\SetWidth{0.9}
\DashLine(-45,-22)(-22,-22){4}
\DashLine(45,-22)(22,-22){4}
\DashLine(-22,-22)(22,-22){4}
\CArc(0,12)(22,0,180)
\Line(-22,12)(22,12)
\Photon(-22,-22)(-22,12){2}{4.5}
\Photon(22,-22)(22,12){-2}{4.5}
\Text(0,-32)[]{$\propV_{SVVFF}$}
\end{picture}
\hspace{-0.3in}
\begin{picture}(98,82)(-49,-39)
\SetWidth{0.9}
\DashLine(-45,-22)(0,-22){4}
\DashLine(45,-22)(0,-22){4}
\CArc(0,12)(22,0,180)
\Line(-22,12)(22,12)
\Photon(0,-22)(-22,12){2}{4.5}
\Photon(0,-22)(22,12){-2}{4.5}
\Text(0,-32)[]{$\propW_{VVFF}$}
\end{picture}
\hspace{-0.3in}
\begin{picture}(98,82)(-49,-39)
\SetWidth{0.9}
\DashLine(-45,-22)(-22,-22){4}
\DashLine(45,-22)(22,-22){4}
\DashLine(-22,-22)(22,-22){4}
\DashCArc(0,12)(22,0,180){4}
\DashLine(-22,12)(22,12){4}
\Photon(-22,-22)(-22,12){2}{4.5}
\Photon(22,-22)(22,12){-2}{4.5}
\Text(0,-32)[]{$\propV_{SVVSS}$}
\end{picture}
\hspace{-0.3in}
\begin{picture}(98,82)(-49,-39)
\SetWidth{0.9}
\DashLine(-45,-22)(0,-22){4}
\DashLine(45,-22)(0,-22){4}
\DashCArc(0,12)(22,0,180){4}
\DashLine(-22,12)(22,12){4}
\Photon(0,-22)(-22,12){2}{4.5}
\Photon(0,-22)(22,12){-2}{4.5}
\Text(0,-32)[]{$\propW_{VVSS}$}
\end{picture}
\hspace{-0.3in}
\begin{picture}(98,82)(-49,-39)
\SetWidth{0.9}
\DashLine(-45,-22)(-20,-22){4}
\DashLine(45,-22)(20,-22){4}
\DashLine(-20,-22)(20,-22){4}
\Photon(-20,-22)(0,7){2}{4.5}
\Photon(20,-22)(0,7){-2}{4.5}
\DashCArc(0,22)(15,-90,270){4}
\Text(0,-32)[]{$\propY_{SVVS}$}
\end{picture}
\hspace{-0.3in}
\begin{picture}(98,82)(-49,-39)
\SetWidth{0.9}
\DashLine(-45,-22)(0,-22){4}
\DashLine(45,-22)(0,-22){4}
\PhotonArc(0,-7)(15,-90,90){-1.8}{5.5}
\PhotonArc(0,-7)(15,90,270){-1.8}{5.5}
\DashCArc(0,23)(15,-90,270){4}
\Text(0,-32)[]{$\propX_{VVS}$}
\end{picture}
\caption{\label{fig:alldiagrams}
The one-loop and two-loop Feynman diagrams for scalar boson
self-energies in the approximation of this paper. Dashed lines
stand for scalars, solid lines for fermions, and wavy lines for massless
vector bosons. Diagrams involving vector boson loops also include the
corresponding ghost loop diagrams. The label for each diagram refers to a
corresponding function obtained as the result of the two-loop
integration. All counterterm diagrams for each diagram are included in
these functions, rendering them ultraviolet finite.  For each diagram with
a fermion loop, fermion mass insertions (indicated by adding a bar to the
corresponding subscript $F$ in the name) are to be made
in all possible ways.  Diagrams indicated by $(*)$ vanish 
identically in
the $\MSbar$ scheme, but not in the $\DRbar$ scheme with non-zero
epsilon-scalar masses.
Ref.~\cite{Martin:2003it} explains the naming
convention.}
\end{figure*}
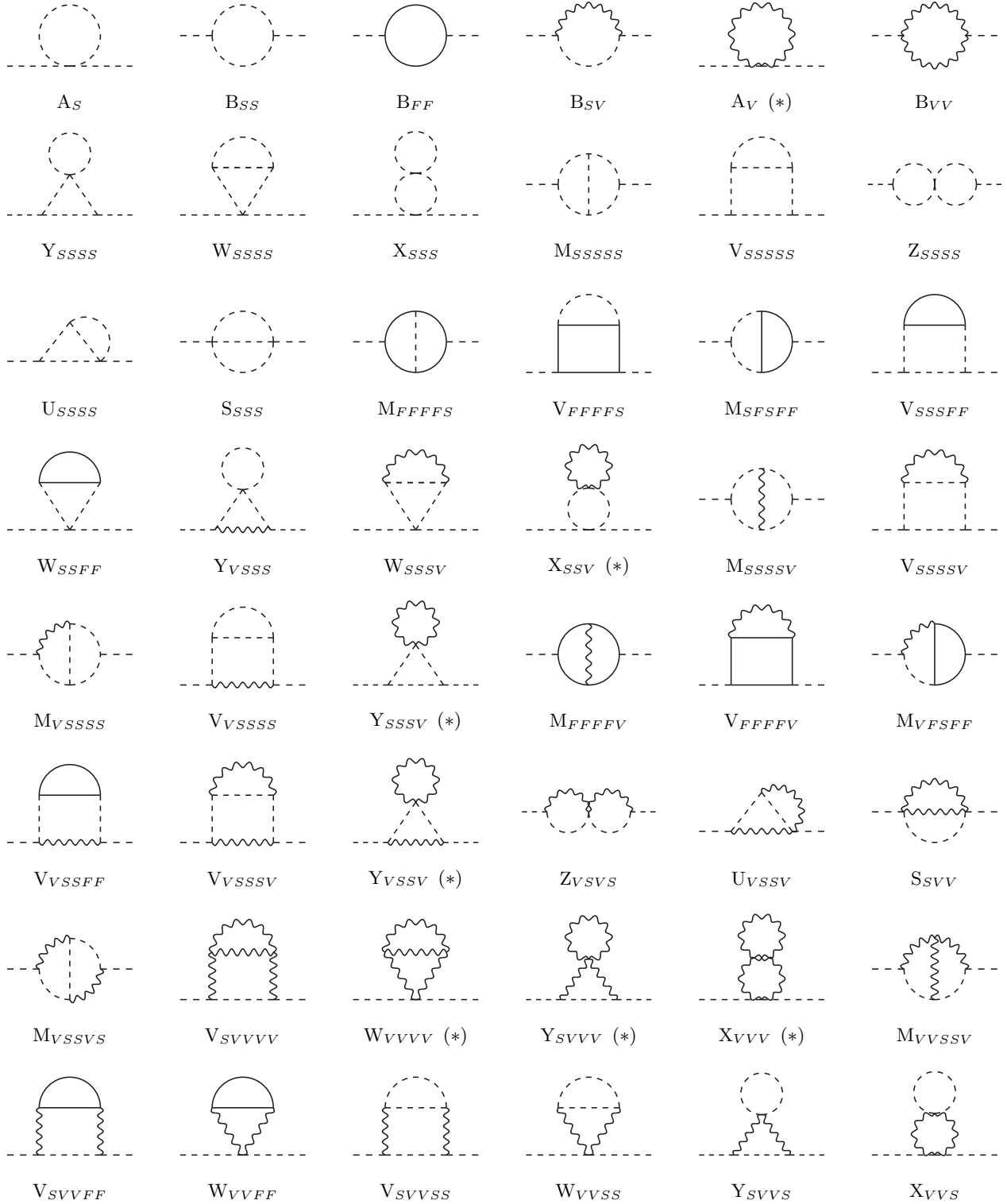


\phantom{.}

\section{Two-loop self-energy functions\label{sec:selfenergy}}
\setcounter{equation}{0}
\setcounter{footnote}{1}

I begin by reviewing the result at one-loop order. Using the notation
for loop integral functions used in \cite{Martin:2003it}, the one-loop
self-energy function matrix for real scalars is:
\begin{widetext}
\beq
\Pi^{(1)}_{ij} &=& 
\frac{1}{2} \lambda^{ijkk} \propA_S (m_k^2) 
+ \frac{1}{2} \lambda^{ikn} \lambda^{jkn} \propB_{SS} (m_k^2, m_n^2)
+ {\rm Re}[y^{KNi} y_{KNj}] \propB_{FF} (m_K^2, m_N^2)
\nonumber \\ &&
+ {\rm Re}[y^{KNi} y^{K'N'j} M_{KK'} M_{NN'} ] \propB_{\Fbar\Fbar} 
  (m_K^2, m_N^2)
+ \delta_{ij} g_a^2 C_a(i) [\propB_{SV}(m_i^2, 0) + \propA_V (0) ]
\eeq
Here I have not written the contribution from massive vector bosons,
which can of course be consistently included in the one-loop part
even if it is neglected in the two-loop part as below.
In ref.~\cite{Martin:2003it}, the results were given in the $\MSbar$
and $\DRbarprime$ schemes, for which $\propA_V(0)$ vanishes. 
In the $\DRbar$ scheme, for massless vector
bosons, one has instead:
\beq
\propA_V(0) = - 2 m_\epsilon^2 .
\eeq
Going from the $\DRbar$ scheme to the $\DRbarprime$ scheme at 
one loop order just removes this term \cite{DRbarprime}. Similarly, at 
two-loop order, the change of
scheme given in \cite{effpot}
removes all terms that depend explicitly on the unphysical
epsilon-scalar mass.
The other functions, due to the first four diagrams in 
fig.~\ref{fig:alldiagrams},
are given in terms of the basis functions by:
\beq
\propA_S(x) &=& A(x) \,\equiv\, x (\lnbar x -1),
\\
-\propB_{SS}(x,y) &=& \frac{1}{2} \propB_{\Fbar\Fbar}(x,y) \,=\, B(x,y)
\equiv  -\int_0^1 dt \, \lnbar[ t x + (1-t) y - t (1-t) s - i \varepsilon]
\\ 
\propB_{FF}(x,y) &=& (x+y-s) B(x,y) - A(x) - A(y)
\\ 
\propB_{SV}(x,0) &=& (3 - \xi) (x+s) B(0,x) + (3 - 2\xi) A(x) + 2 (\xi-1) s
\\
&=& 4s +3 x - 3 x \lnbar x 
+ (\xi-3) (s -x) [\lnbar x + (1 +x/s) \ln (1-s/x- i\varepsilon)] .
\eeq

At two-loop order, one can write the self-energy function as a sum of
contributions from diagrams with 0, 1, and 2 or more vector lines:
\beq
\Pi^{(2)}_{ij} = \Pi^{(2,0)}_{ij} + \Pi^{(2,1)}_{ij} + \Pi^{(2,2)}_{ij} .
\eeq

First, consider $\Pi^{(2,2)}_{ij}$, the two-loop self-energy contributions
from diagrams with two or more massless vector boson (or ghost) propagators.
The pertinent individual diagrams shown in fig.~\ref{fig:alldiagrams}
contribute with group theory factors proportional
to:
\beq
\propV_{VSSSV}, \propY_{VSSV} 
:
&&
g_a^2 g_b^2 C_a(i) C_b(i) 
\\
\propZ_{VSVS}, \propU_{VSSV}, \propS_{SVV} 
:
&&
g_a^2 C_a(i) [ g_b^2 C_b(i) - \frac{1}{4} g_a^2 C_a(G)] 
\\
\propM_{VSSVS}
: 
&&
g_a^2 C_a(i) [ g_b^2 C_b(i) - \frac{1}{2} g_a^2 C_a(G)] 
\phantom{xxxxxxxxxxx}
\\
\propV_{SVVVV}, \propW_{VVVV}, \propY_{SVVV}, \propX_{VVV} ,
\propM_{VVSSV}
:
&&
g_a^4 C_a(i) C_a(G) 
\\
\propV_{SVVFF}, \propV_{SVV\Fbar\Fbar}, \propW_{VVFF}, \propW_{VV\Fbar\Fbar}
:
&&
g_a^4 C_a(i) I_a(R_{\rm fermions})
\\
\propV_{SVVSS}, 
\propW_{VVSS}, 
\propY_{SVVS}, 
\propX_{VVS}
:
&&
g_a^4 C_a(i) I_a(R_{\rm scalars})
.
\eeq
Reorganizing the contributions in terms of the four independent
group theory factors,
and writing them in terms of the basis functions, I obtain: 
\beq
\Pi^{(2,2)}_{ij} &=& \delta_{ij} g_a^2 C_a(i) \Bigl [
g_b^2 C_b(i) F_1(m^2_i)
+ g_a^2 C_a(G) F_2(m^2_i)
+ g_a^2 I_a(K) F_3(m^2_i,m_K^2)
+ g_a^2 I_a(k) F_4(m^2_i,m_k^2)
\Bigr ]\phantom{xxx}
\eeq
with loop integral functions defined by:
\beq
&& F_1(x) \>=\>
- (3x+s)^2 M(0,x,x,0,x)
-(12 x + 4 s) U(x,0,x,x)
+ 8x U(x,0,0,0) 
+ (2 s - 6 x) T(x,0,0)
\nonumber \\ && \qquad\>\>
-3 S(x,x,x)
+ [(3x+s) B(0,x) + 4s - 12 x + (6 + 2s/x) A(x)] B(0,x) + 7 A(x)^2/x 
\nonumber \\ && \qquad\>\>
- 19 A(x) + 10 x + 15 s/2
+ 4 [4 x - 3 A(x) ] (x/s) \ln(1-s/x) 
\nonumber \\ && \qquad\>\>
+(1-\xi)  \Bigl \lbrace -8x U(x,0,0,0) + 4 (s+x) T(x,0,0)
+ 4 [x B(0,x) + (1+s/x) A(x) + 3 x + s] B(0,x)
\nonumber \\ && \qquad\>\>
+ A(x)^2/x - 5 A(x) + 22x + 2 s
+ 2 [4 x - 3 A(x) ] (x/s) \ln(1-s/x)
\Bigr\rbrace
\nonumber \\ && \qquad\>\>
+(1-\xi)^2  \Bigl \lbrace
-2x U(x,0,0,0)
+ (s+x) T(x,0,0)
+ [x B(0,x) +(1+s/x) A(x) + 4 x + 2 s] B(0,x)
\nonumber \\ && \qquad\>\>
+ A(x)^2/x + 2 A(x) + 5x/2 -3s/2
\Bigr\rbrace
+ \delta_{\overline{\rm MS}} [s - 2 x + 4 A(x)]
\nonumber \\ && \qquad\>\>
+ m^2_{\epsilon} \Bigl [
2 A(x)/x - 2 - (8x/s) \ln(1-s/x) 
+ (1-\xi) \lbrace -2 A(x)/x - 6 - (4x/s) \ln(1-s/x) \rbrace \Bigr ]
\label{eq:defineFone}
\\
&& F_2(x) \>=\> 
[(3 x + s)^2/2] M(0,x,x,0,x) + (s-x)^2 M(0,0,x,x,0)
+ (6 x + 2 s) U(x,0,x,x) 
+ 10 x U(x,0,0,0)
\nonumber \\ && \qquad\>\>
+ (6s-10x) T(x,0,0) + (3/2) S(x,x,x)
+ (13 s/2 - x/2) B(0,x)^2 
\nonumber \\ && \qquad\>\>
+ [7s/3 - 77x/3 + 6 (1 + s/x) A(x)] B(0,x) 
+ 6 A(x)^2/x 
- 47 A(x)/3 -10 x + 49 s/4
\nonumber \\ && \qquad\>\>
+ (1-\xi) \Bigl \lbrace 
[(s-x)^2/2] M(0,0,x,x,0) + (s/2-11x/2) T(x,0,0) + 5x U(x,0,0,0)
\nonumber \\ && \qquad\>\>
+ [(3 s-x) B(0,x) + (3 + s/x) A(x) - 5s - 29x] B(0,x)/2 
+ A(x)^2/x - 7 A(x) -8x + 4s
\Bigr \rbrace
\nonumber \\ && \qquad\>\>
+ (1-\xi)^2 \Bigl \lbrace 
x T(x,0,0) 
-(s/2 + x/2) U(x,0,0,0)
+ [9x/4 + 5s/4] B(0,x)
+ 3 A(x)/2 + 5x/8 -s/8
\Bigr \rbrace
\nonumber \\ && \qquad\>\>
+ \delta_{\overline{\rm MS}}
[(s+x) B(0,x) + x/2]
- 10 m_{\epsilon}^2
\\
&& F_3(x,y) \>=\> 
\lbrace -4 x [(s-x)^2 + 16 (s+x) y - 24 y^2]/15 y (s-x) \rbrace T(x,y,y)
\nonumber \\ && \qquad\qquad
+ 
\lbrace
[22 s^3 - 14 s^2 x - 38 s x^2 + 30 x^3 - 120 s^2 y + 48 s x y + 8 x^2 y +
 128 s y^2 - 64 x y^2]/15 (s-x)^2 \rbrace T(y,y,x)
\nonumber \\ && \qquad\qquad
+
\lbrace 
[12 s^2 x - 4 s^3 - 12 s x^2 + 4 x^3 - 86 s^2 y + 44 s x y + 42 x^2 y +
 128 s y^2 - 64 x y^2]/15 y (s-x)^2 \rbrace S(x,y,y)
\nonumber \\ && \qquad\qquad
+ \lbrace 2 (s+x) [A(y)/y+1] + 2 (s-x)^2/15 y \rbrace B(0,x)
+ \Bigl \lbrace
[8 (s-x)^2 + 32 (s+x) y] A(x) A(y) 
\nonumber \\ && \qquad\qquad
+ [4 s x - 26 s^2 + 22 x^2  + 64 s y - 32 x y] A(y)^2
+ [2 s^3 - 6 s^2 x + 6 s x^2 - 2 x^3 + 52 s^2 y - 40 s x y 
- 12 x^2 y 
\nonumber \\ && \qquad\qquad
-32 s y^2 + 64 x y^2] A(x)
+ [56 s^2 y - 8 x (s-x)^2 + 144 s x y - 264 x^2 y - 256 s y^2 
+ 128 x y^2] A(y) 
\nonumber \\ && \qquad\qquad
+ 5 s^4/2 - 23 s^3 x/2 + 39 s^2 x^2/2 - 29 s x^3/2 + 4 x^4 + 
20 s^3 y - 38 s^2 x y - 32 s x^2 y + 50 x^3 y 
\nonumber \\ && \qquad\qquad
- 244 s^2 y^2 + 240 s x y^2 
-100 x^2 y^2 + 256 s y^3 - 128 x y^3
\Bigr \rbrace/15 y (s-x)^2
+ \deltaMSbar (-2 y) + 2 m_{\epsilon}^2
\\
&& F_4(x,y) \>=\> 
\lbrace -x [ (s-x)^2 + 56 (s+x) y + 96 y^2 ]/15 y (s-x) \rbrace T(x,y,y)
\nonumber \\ && \qquad\qquad
+
\lbrace [13 s^3 - 11 s^2 x - 20 s^2 y 
- 17 s x^2 - 8 s x y - 128 s y^2 
+ 15 x^3  + 92 x^2 y + 64 x y^2]/15 (s-x)^2\rbrace T(y,y,x)
\nonumber \\ && \qquad\qquad
+ 
\lbrace [(x-s)^3 -69 s^2 y + 26 s x y + 43 x^2 y - 128 s y^2 + 64 x y^2]/
15 y (s-x)^2 \rbrace S(x,y,y)
\nonumber \\ && \qquad\qquad
+ \lbrace (s+x)[A(y)/y+1] + (s-x)^2/30 y \rbrace B(0,x)
+ \Bigl \lbrace
[ 2 (s-x)^2  - 32 (s + x)  y] A(x) A(y)
\nonumber \\ && \qquad\qquad
+ [ 32 x y -14 s^2 - 44 s x + 58 x^2  - 64 s y] A(y)^2
+ [(s-x)^3/2 + 28 y (s^2 - x^2) + 32 y^2 (s - 2x)] A(x)
\nonumber \\ && \qquad\qquad
+ [4 s x^2 - 2 s^2 x - 2 x^3 + 164 s^2 y - 104 s x y + 4 x^2 y 
+ 256 s y^2 - 128 x y^2 ] A(y)
+ 5 s^4/8 - 23 s^3 x/8 
\nonumber \\ && \qquad\qquad
+ 39 s^2 x^2/8 
- 29 s x^3/8 + x^4 
+ 45 s^3 y/2 - 42 s^2 x y - 51 s x^2 y/2 
+ 45 x^3 y + 24 s^2 y^2 -280s x y^2 
\nonumber \\ && \qquad\qquad
+ 360 x^2 y^2 
- 256 s y^3 + 128 x y^3
\bigr \rbrace/ 15 y (s-x)^2
+  \delta_{\overline{\rm MS}} 4 A(y)
.
\label{eq:defineFfour}
\eeq
In the limit $s \rightarrow x$,
the function $F_1$ has a logarithmic singularity 
of the form $\ln (1-s/x)$, and $F_3$ and $F_4$ 
have $1/(s-x)$ and $1/(s-x)^2$
singularities in individual terms. After taking into account
the identities mentioned in Appendix A, one can check that $F_2$, $F_3$,
and $F_4$ are actually finite in that limit. 
[Because of these same identities, which 
hold between
the basis functions due to coincident and vanishing arguments, the 
representations given in 
eqs.~(\ref{eq:defineFone})-(\ref{eq:defineFfour})
are not unique.]
Also, the functions $F_1$ and $F_2$ are both 
dependent on the gauge-fixing parameter $\xi$. 
It is therefore useful to define, for the 
purpose of computing the finite and 
gauge-invariant pole mass, the functions:
\beq
\widetilde F_1(x) 
&\equiv& 
\lim_{s \rightarrow x}
\left [
F_1(x) + 
[\propB_{SV}(x,0) + \propA_V(0)] 
\propB'_{SV}(x,0) 
\right ]
\nonumber \\ 
&=&
x \Bigl (12 \pi^2 - 16 \pi^2 \ln 2 
- \frac{11}{8} + 24 \zeta(3)
- \frac{39}{2} \lnbar x + \frac{15}{2} \lnbar^2 x 
\Bigr )
+ \deltaMSbar x (4 \lnbar x -5) + m^2_{\epsilon} (6 \lnbar x - 4)
\label{eq:F1tilde}
\\
\widetilde F_2(x)  
&\equiv& 
\lim_{s \rightarrow x}
F_2(x) 
\nonumber \\
&=& x \Bigl (
\frac{1147}{16} 
- \frac{10 \pi^2}{3} 
+ 8 \pi^2 \ln 2 
- 12 \zeta(3) 
- \frac{409}{12}\lnbar x 
+ \frac{19}{4} \lnbar^2 x
\Bigr ) 
+ \deltaMSbar x\left (\frac{9}{2} - 2 \lnbar x\right ) 
- 10 m^2_{\epsilon}
\\
\widetilde F_3(x,y) 
&\equiv& 
\lim_{s \rightarrow x} F_3(x,y) 
\nonumber \\ 
&=& 
2 (x-y^2/x) {\rm Li}_2(1-x/y) 
+ 8 (x-y) f (\sqrt{y/x})
+ y \Bigl ( 18 + \frac{\pi^2 y}{3 x}  - 6 \lnbar x  - 6 \lnbar y \Bigr )
\nonumber \\ &&
+ x \Bigl (
-\frac{49}{4} - \frac{\pi^2}{3} + \frac{19}{3} \lnbar x  -2 \lnbar x 
\lnbar y + \lnbar^2 y \Bigr )
+ \deltaMSbar (-2y) + 2 m_{\epsilon}^2 
\\
\widetilde F_4(x,y) 
&\equiv& 
\lim_{s \rightarrow x} F_4(x,y) 
\nonumber \\
&=& 
(x + 6 y + y^2/x) {\rm Li}_2(1-x/y) + 8 (x + y) f(\sqrt{y/x})
+ y [-4 - \pi^2 (1 + y/6x) + 7 \ln(x/y)
+ 3 \lnbar^2 y ]
\nonumber \\ &&
+ x \Bigl (-\frac{75}{8} - \frac{\pi^2}{6} + \frac{25}{6} \lnbar x 
- \lnbar x \lnbar y 
+ \frac{1}{2} \lnbar^2 y \Bigr ) + \deltaMSbar 4 y (\lnbar y - 1)
\label{eq:F4tilde}
\eeq
where
\beq
f(z) &=& 
z \lbrace {\rm Li}_2 ([1-z]/[1+z])
- {\rm Li}_2 ([z-1]/[1+z]) 
+ \pi^2/4 \rbrace
.
\label{eq:definefunky}
\eeq
\end{widetext}
The reason for including the term proportional to 
$\propB'_{SV}(x,0) \equiv
\partial \propB_{SV}(x,0)/\partial s$ in the definition of 
$\widetilde F_1$ in eq.~(\ref{eq:F1tilde})
is that this includes the appropriate contribution from 
the $\Pi^{(1)} \Pi^{(1)\prime}$ term in the formula for the pole mass,
eq.~(\ref{eq:poleexp}), exhibiting the 
simultaneous cancellation of the gauge dependence and the 
logarithmic singularity in the limit $s\rightarrow x$ .

Some useful limits are, for $x=y$:
\beq
\widetilde F_3(x,x) &=& x \Bigl (
\frac{23}{4} - \frac{17}{3} \lnbar x - \lnbar^2x
\Bigr ) 
\nonumber \\ &&
+ \deltaMSbar (-2x)
+ 2 m^2_{\epsilon}
,
\\
\widetilde F_4(x,x) &=& x \Bigl (
\frac{8 \pi^2}{3} - \frac{107}{8} + \frac{25}{6} \lnbar x
+ \frac{5}{2} \lnbar^2x \Bigr ) 
\nonumber \\ &&
+ \deltaMSbar 4 x (\lnbar x-1)
,
\eeq
and for $y=0$:
\beq
\widetilde F_3(x,0) &=& x \Bigl (
\frac{19}{3} \lnbar x
- \lnbar^2x 
-\frac{49}{4} 
- \frac{2 \pi^2}{3} 
\Bigr )
+ 2 m_{\epsilon}^2 
,
\phantom{xxxx}
\\
\widetilde F_4(x,0) &=& x \Bigl (
\frac{25}{6} \lnbar x
- \frac{1}{2} \lnbar^2 x 
-\frac{75}{8} 
- \frac{\pi^2}{3} 
\Bigr )
,
\phantom{xxxx}
\eeq 
and for $x=0$:
\beq
\widetilde F_3(0,y) &=& y (4 - 12 \lnbar y) + \deltaMSbar (-2y) +
2 m_\epsilon^2
,
\phantom{xxxx}
\label{eq:Fthreeoy}
\\
\widetilde F_4(0,y) &=& y (11 + 3 \lnbar^2 y) + \deltaMSbar 4y(\lnbar y-1)
.
\label{eq:Ffouroy}
\eeq

Next, consider $\Pi^{(2,1)}$, the contributions from diagrams with one 
vector line. 
These were already given in sections IV.C, IV.D, IV.E and V 
of ref.~\cite{Martin:2003it}, but the results can be rewritten in a 
somewhat nicer form in the case of massless vector bosons:
\begin{widetext}
\beq
\Pi^{(2,1)}_{ij} &=& 
\frac{1}{2} \lambda^{ijkk} g_a^2 \Bigl \lbrace
C_a(k) G_S (m_k^2) + C_a(i) G_{SSS} (m_i^2, m_j^2, m_k^2)
\Bigr\rbrace 
\nonumber \\
&& 
+ \frac{1}{2} \lambda^{ikn} \lambda^{jkn} g^2_a 
\Bigl \lbrace 
\frac{1}{2} [C_a(k) + C_a(n) - C_a(i)] G_{SS} (m_k^2, m_n^2)
\nonumber \\
&& 
+ [C_a(i) - C_a(k) + C_a(n)] G_{SSSS} (m_i^2, m_j^2, m_k^2, m_n^2)
\Bigr \rbrace
\nonumber \\
&& 
+ {\rm Re}[y^{KNi} y_{KNj}] g_a^2 \Bigl \lbrace 
\frac{1}{2} [C_a(K) + C_a(N) - C_a(i)] G_{FF} (m_K^2, m_N^2)
\nonumber \\
&& 
+ [C_a (i) - C_a (K) + C_a (N)] G_{SSFF} (m_i^2, m_j^2, m_K^2, m_N^2)
\Bigr \rbrace
\nonumber \\
&& 
+ {\rm Re}[y^{KNi} y^{K'N'j} M_{KK'} M_{NN'}] g_a^2 \Bigl \lbrace 
\frac{1}{2}[C_a(K) + C_a(N) - C_a(i)] G_{\Fbar\Fbar} (m_K^2, m_N^2)
\nonumber \\ 
&& 
+ [C_a (i) - C_a (K) + C_a (N)] 
G_{SS\Fbar\Fbar} (m_i^2, m_j^2, m_K^2, m_N^2)
\Bigr \rbrace
\eeq
The functions 
$G_{SS}$, $G_{FF}$, $G_{\Fbar\Fbar}$,
$G_{SSSS}$, 
$G_{SSFF}$, 
$G_{SS\Fbar\Fbar}$, were defined in equations (5.31)-(5.36) of
\cite{Martin:2003it}, referring to earlier results in that paper.
Also I define here:
\beq
G_S(y) &=& \propW_{SSSV} (y,y,y,0) + \propX_{SSV} (y,y,0) , 
\\
       &=& y (-12 + 11 \lnbar y - 3 \lnbar^2 y) 
           - 2 m^2_{\epsilon} \lnbar y 
\\
G_{SSS} (x,y,z) &=& \propY_{VSSS}(0,x,y,z) 
\\
                &=& A(z) [(3 - 2 \xi) A(x) + (s + x) (3 - \xi) B(0, x)]/(x-y)
                    + (x \leftrightarrow y),
\eeq
in terms of functions appearing in eqs.~(5.1), (5.2), (5.4), and (5.5)
of ref.~\cite{Martin:2003it}. All of these functions 
were written in that paper
in the $\MSbar$ and $\DRbarprime$ schemes in a notation consistent with
eq.~(\ref{eq:deltaMSbar}) 
of the present paper. To obtain the $\DRbar$ scheme results,
one should add $-2 m_\epsilon^2 \lnbar x$ to eq.~(5.2), and
$2 m_\epsilon^2 B(x,y')$ to eq.~(5.3), and 
$-2 m_\epsilon^2$ to eq.~(5.17), all in ref.~\cite{Martin:2003it}.

The functions $G_S(x)$, $G_{SS}(x,y)$, $G_{FF} (x,y)$, and 
$G_{\Fbar\Fbar}(x,y)$
are each independent of the gauge-fixing parameter $\xi$. The other
functions are are not. However, for the purpose of computing the
two-loop pole mass, one can define the gauge-invariant functions:
\beq
\widetilde G_{SSS} (x,x,z) &=& \lim_{s\rightarrow x} \left [
G_{SSS} (x,x,z) + \propA_{S}(z) 
\propB'_{SV}(x,0)
\right ]
,
\label{eq:GSSStilde}
\\
\widetilde G_{SSSS} (x,x,y,z) &=& \lim_{s\rightarrow x} \left [
G_{SSSS} (x,x,y,z) + 
\propB_{SS}(y,z) 
\propB'_{SV}(x,0)
\right ]
,
\\
\widetilde G_{SSFF} (x,x,y,z) &=& \lim_{s\rightarrow x} \left [
G_{SSFF} (x,x,y,z) + 
\propB_{FF}(y,z) 
\propB'_{SV}(x,0)
\right ]
,
\\
\widetilde G_{SS\Fbar\Fbar} (x,x,y,z) &=& \lim_{s\rightarrow x} \left [
G_{SS\Fbar\Fbar} (x,x,y,z) + 
\propB_{\Fbar\Fbar}(y,z) 
\propB'_{SV}(x,0)
\right ]
.
\label{eq:GSSfftilde}
\eeq
The first two arguments are taken equal in these functions, 
because one can consistently neglect the off-diagonal entries
in the two-loop part of the self-energy when computing the two-loop 
pole mass. As before, the reason for including the terms involving 
$\propB'_{SV}(x,0) = \partial \propB_{SV}(x,0)/\partial s$ 
is that this naturally includes
the corresponding parts of the term involving 
$\Pi^{(1)} \Pi^{(1)\prime}$ in 
eq.~(\ref{eq:poleexp}).

The result of taking the limits in 
eqs.~(\ref{eq:GSSStilde})-(\ref{eq:GSSfftilde}) is:
\beq
\widetilde G_{SSS}(x,x,z) &=& (4 - 3 \lnbar x) A(z)
\label{eq:GSSStildeexp}
\\
\widetilde G_{SSSS}(x,x,y,z) &=& 
4 (x - y + z) M(0, z, x, y, z) 
+ 4 U(x, 0, z, z)
+ 2 U(z, y, z, x)
+ 4 {\overline T} (0, y, z)
- 4 S(0,y,z)/x 
\nonumber \\ &&
- 2 y T(y,0,z)/x 
+ (2 - 2z/x) T (z,0,y)
+ 2 I(x,y,z)/x 
+ 3 I(x',y,z) 
+ z (3 \lnbar z -7) B(y,z')
\nonumber \\ &&
+ 2 (\lnbar x + \lnbar z - 8) B(y,z)
+ (2 y \lnbar y + 2 z \lnbar z -7x/2 -4y -4z)/x
\\
\widetilde G_{SS\Fbar\Fbar}(x,x,y,z) &=& - 2 G_{SSSS}(x,x,y,z)
+ 6 z B(y,z') (1 - \lnbar z) + \deltaMSbar [-4 z B(y,z')]
\\
\widetilde G_{SSFF}(x,x,y,z) &=&
4[(x-y)^2 - z^2] M (0,z,y,x,z) 
+ 4 (x - y - z) U(x, 0, z, z) 
- 4 y U(z, y, z, x)
\nonumber \\ &&
+ 4 (x - y - z) {\overline T}(0, y, z)
+ (x - 2 z) (x - y - z) T(z, 0, y)/x
+ 2 y (y + z) T(y, 0, z)/x
\nonumber \\ &&
+ (4 (y+z)/x  -1) S(0, y, z)
+ 2 S(x, z, z)
+ 3 (x - y - z) I (x', y, z)
\nonumber \\ &&
+ [1 -2(y+z)/x] I(x,y,z)
+ 2 (x - y - z) z (3 \lnbar z - 5) B(y, z')
+
[15 y + 17 z  - 7x 
\nonumber \\ &&
- 2 (x + y + z) \lnbar x + (x - y - 3 z) \lnbar z]  B(y,z)
+ 3 \lnbar x (y \lnbar y - y + z \lnbar z -z) 
\nonumber \\ &&
- y [3 + 2 (y+z)/x] \lnbar y  
+ z [5 \lnbar z -15 - 2 (y+z)/x] \lnbar z
+ 4 (y+z)^2/x -15 x/4  + 11 y/2 + 33 z/2
\nonumber \\ &&
+ \deltaMSbar [(y - x - z) B(y, z) + 2 (x - y - z) z B(y, z')
- y \lnbar y + z \lnbar z -3x/4 + 2y + 2z ]
\label{eq:GSSFFtildeexp}
\eeq

Finally, consider the contributions from diagrams without any gauge
interactions, $\Pi^{(2,0)}_{ij}$. These were
already given in sections IV.A and IV.B of
ref.~\cite{Martin:2003it} in exactly the same notation, 
and so will not be repeated here. However, for the purpose of computing
the pole mass, it is convenient to define:
\beq
\widetilde \Pi^{(2,0)}_i &=&
\lim_{s \rightarrow m_{i}^2} \Bigl [
\Pi^{(2,0)}_{ii}
+ \Pi^{(1)}_{ii} \Bigl \lbrace
   \frac{1}{2} (\lambda^{ikn})^2 \propB'_{SS} (m_k^2, m_n^2)
  + |y^{KNi}|^2
  \propB_{FF}' (m_K^2, m_N^2)
\nonumber \\ &&
  + {\rm Re}[y^{KNi} y^{K'N'i} M_{KK'} M_{NN'}]
  \propB_{\Fbar\Fbar}' (m_K^2, m_N^2)\Bigr\rbrace \Bigr ],
  \phantom{xx} 
\label{eq:ide}
\eeq
where the primes indicate differentiation with respect to $s$. This
incorporates the rest of the terms involving $\Pi^{(1)}
\Pi^{(1)\prime}$ in eq.~(\ref{eq:poleexp}). (The derivatives of
one-loop functions with respect to $s$ are easy to obtain analytically,
and are given in the present notation in
ref.~\cite{Martin:2003it}.)

The pole mass is now obtained as:
\beq
M^2_{i} &=& m^2_{i} 
+ \frac{1}{16 \pi^2} \widetilde \Pi^{(1)}_{ii} 
+ \frac{1}{(16 \pi^2)^2} \Bigl [
\widetilde \Pi^{(2,0)}_{i} 
+ \widetilde \Pi^{(2,1)}_{i} 
+ \widetilde \Pi^{(2,2)}_{i} 
+ \sum_{j\not= i} (\widetilde \Pi^{(1)}_{ij})^2/
     (m_{i}^2 - m_{j}^2) \Bigr ],
\label{eq:poleresultgen}
\eeq
\end{widetext}
with no sum on $i$ implied. Here $\widetilde \Pi^{(2,1)}_i$ is obtained from
$\Pi^{(2,1)}_{ii}$ by replacing the functions
$G_{SSS}$, $G_{SSFF}$ and $G_{SS\Fbar\Fbar}$ with 
$\widetilde G_{SSS}$,
$\widetilde G_{SSFF}$, and
$\widetilde G_{SS\Fbar\Fbar}$, and 
$\widetilde \Pi^{(2,2)}_{i}$ is obtained from 
$\Pi^{(2,2)}_{ii}$ by replacing the functions 
$F_{1,2,3,4}$ with $\widetilde F_{1,2,3,4}$, and
\beq
\widetilde \Pi^{(1)}_{ij} = \lim_{s \rightarrow m^2_i} 
\Pi^{(1)}_{ij}.
\eeq 
It is a nice check that the limit $s\rightarrow x$ gives
a finite result for the pole mass here. 
Also, the
independence of the pole mass with respect to the choice of
gauge-fixing parameter, 
up to terms of three-loop order, now follows immediately 
from the absence of $\xi$ in
eqs.~(\ref{eq:F1tilde})-(\ref{eq:F4tilde})
and (\ref{eq:GSSStildeexp})-(\ref{eq:ide}).
Note that this relies on 
cancellations involving the
two-loop and iterated one-loop
self-energy function contributions to the pole mass.

\section{Two-loop SUSYQCD corrections to squark 
self-energies and pole masses\label{sec:squarks}}
\setcounter{equation}{0}
\setcounter{footnote}{1}

As an example application of the preceding results, consider the two-loop
strong (SUSYQCD) contributions to the squark masses in supersymmetry.
Consider an approximation in which the squark mixings respect
family symmetry, but can mix left- and right-handed squarks. (This is
a slight generalization of the usual assumption that only the third-family
squarks have
a significant mixing.) The tree-level 
squared-mass matrices for each squark flavor
can then each be treated as $2\times 2$ matrices ${\bf m}_{\tilde q}^2$
in the gauge eigenstate basis. 
They are diagonalized by unitary transformations:
\beq
\begin{pmatrix} \tilde q_L \\ \tilde q_R \end{pmatrix}
= X_{\tilde q}
\begin{pmatrix} \tilde q_1 \\ \tilde q_2 \end{pmatrix} 
,
\eeq
for $q = u,d,c,s,t,b$, chosen so that
\beq
X_{\tilde q}^{-1} {\bf m}^2_{\tilde q} X_{\tilde q} =
\begin{pmatrix} m_{\tilde q_1}^2 & 0 \\
         0 & m^2_{\tilde q_2} \end{pmatrix} .
\label{eq:diagsquarks}
\eeq
Thus, $\tilde q_1$, $\tilde q_2$ are mass eigenstates with squared masses
$m^2_{\tilde q_1}$, $m^2_{\tilde q_2}$, while $\tilde q_L$, $\tilde q_R$
are the gauge eigenstates.
Unitarity of the matrix $X_{\tilde q}$ allows one to write 
\beq
X_{\tilde q} = \begin{pmatrix} L_{\tilde q_1} & L_{\tilde q_2} \\
            R_{\tilde q_1} & R_{\tilde q_2} \end{pmatrix},
\eeq
where 
$L_{\tilde q_1} = R_{\tilde q_2}^* = c_{\tilde q}$, and
$R_{\tilde q_1} = -L_{\tilde q_2}^* = s_{\tilde q}$, with
\beq
|c_{\tilde q}|^2 + |s_{\tilde q}|^2 = 1.
\eeq
If the off-diagonal elements of the squared mass matrix are real, then
$c_{\tilde q}$ and $s_{\tilde q}$ are the cosine and sine of a
squark mixing angle for each of $q=t,b$. Also, 
to a good approximation in most realistic models,
$s_{\tilde q}=1$ and $c_{\tilde q} = 0$ for $q = u,d,c,s$.
For convenience, I define the following combinations:
\beq
P_i^j 
&=&
L_{\squark_i} L_{\squark_j}^* - R_{\squark_i} R_{\squark_j}^*,
\\
N_i^j
&=&
L_{\squark_i} R_{\squark_j}^* + R_{\squark_i} L_{\squark_j}^*,
\eeq
when $\squark_i$ and $\squark_j$ are of the same flavor, and
$P_i^j = N_i^j = 0$ otherwise.
(Here, squarks are complex scalars, so the heights of indices
are significant.)
Also, in the following 
$C_q = 4/3$ and $C_G = 3$ are the quadratic Casimir invariants of the 
fundamental and adjoint representations, and $I_q = 1/2$ is the Dynkin
index of the fundamental. 

The self-energies will then likewise be $2\times 2$ matrices
for each of the 6 flavors.
The SUSYQCD contribution
to the one-loop self-energy in the $\DRbarprime$ scheme is:
\begin{widetext}
\beq
\Pi^{(1)j}_i =
g_3^2 C_q \Bigl [
P_i^k P_k^j \propA_{S} (\squark_k)
+ 2 \delta_i^j \propB_{FF}(q,\gluino)
- 2 N_i^j m_q m_{\gluino} \propB_{\Fbar\Fbar}(q,\gluino)
+  \delta_i^j \propB_{SV}(\squark_i,0) 
\Bigr ] ,
\label{eq:squarksoneloopmixed}
\eeq
where $k$ is summed over the two mass eigenstates of the same flavor
as $i,j$.
(The other, non-SUSYQCD, 
one-loop contributions can be found in ref.~\cite{Pierce:1996zz}.)

The contributions from two-loop SUSYQCD 
diagrams with no gluon propagators are:
\beq
\Pi^{(2,0)j}_i &=& 
4 g_3^4 C_q^2 \Bigl [
\frac{1}{4} P_i^k P_k^m P_m^n P_n^j 
   \propX_{SSS} (\squark_k, \squark_n,\squark_m)
+  \frac{1}{2} P_i^k P_k^j 
   \propW_{SSFF} (\squark_k,\squark_k,\quark,\gluino)
\nonumber \\ &&
-  \frac{1}{2} P_i^k N_k^m P_m^j 
   m_{\quark} m_{\gluino}
   \propW_{SS\Fbar\Fbar} (\squark_k,\squark_m,\quark,\gluino)
+  
  (L_{\squark_i} L_{\squark_j}^* |L_{\squark_k}|^2  
  +R_{\squark_i} R_{\squark_j}^* |R_{\squark_k}|^2)
 \propV_{FFFFS} (\gluino,\quark,\quark,\gluino,\squark_k)
\nonumber \\ &&
- \delta_i^j N_k^k
  m_{\quark} m_{\gluino} 
  \propV_{FF\Fbar\Fbar S}
  (\gluino,q,q,\gluino,\squark_k)
+  
  (L_{\squark_i} R_{\squark_j}^* R_{\squark_k} L_{\squark_k}^*
  +R_{\squark_i} L_{\squark_j}^* L_{\squark_k} R_{\squark_k}^*)
  m_{\gluino}^2
  \propV_{\Fbar FF\Fbar S}
  (\gluino,q,q,\gluino,\squark_k)
\nonumber \\ &&
+ 
  (L_{\squark_i} L_{\squark_j}^* |R_{\squark_k}|^2
  +R_{\squark_i} R_{\squark_j}^* |L_{\squark_k}|^2)
  m_q^2
  \propV_{F\Fbar\Fbar FS}
  (\gluino,q,q,\gluino,\squark_k)
-N_i^j 
  m_q m_{\gluino}
  \propV_{\Fbar F\Fbar FS}
  (\gluino,q,q,\gluino,\squark_k)
\nonumber \\ &&
+
  (L_{\squark_i} R_{\squark_j}^* L_{\squark_k} R_{\squark_k}^*
  +R_{\squark_i} L_{\squark_j}^* R_{\squark_k} L_{\squark_k}^*)
  m_q^2 m_{\gluino}^2
  \propV_{\Fbar\Fbar\Fbar\Fbar S}
  (\gluino,q,q,\gluino,\squark_k)
\Bigr ]
\nonumber \\ &&
+ g_3^4 [4 C_q^2 - 2C_G C_q] \Bigl [
 (L_{\squark_i} R_{\squark_j}^* R_{\squark_k} L_{\squark_k}^*
 +R_{\squark_i} L_{\squark_j}^* L_{\squark_k} R_{\squark_k}^*)
 \propM_{FFFFS} (\quark,\gluino,\gluino,\quark,\squark_k)
\nonumber \\ &&
- N_i^j
  m_{\quark} m_{\gluino} 
  \propM_{FF\Fbar\Fbar S}
  (\quark,\gluino,\gluino,\quark,\squark_k)
+ 
  (L_{\squark_i} L_{\squark_j}^* |L_{\squark_k}|^2  
  +R_{\squark_i} R_{\squark_j}^* |R_{\squark_k}|^2)
  m_{\gluino}^2
  \propM_{F\Fbar\Fbar F S}
  (\quark,\gluino,\gluino,\quark,\squark_k)
\nonumber \\ &&
- \delta_i^j N_k^k 
  m_{\quark} m_{\gluino}
  \propM_{F\Fbar F\Fbar S}
  (\quark,\gluino,\gluino,\quark,\squark_k)
+ 
  (L_{\squark_i} L_{\squark_j}^* |R_{\squark_k}|^2  
  +R_{\squark_i} R_{\squark_j}^* |L_{\squark_k}|^2)
  m_{\quark}^2
  \propM_{F\Fbar\Fbar F S}
  (\gluino,\quark,\quark,\gluino,\squark_k)
\nonumber \\ &&
+ 
  (L_{\squark_i} R_{\squark_j}^* L_{\squark_k} R_{\squark_k}^*  
  +R_{\squark_i} L_{\squark_j}^* R_{\squark_k} L_{\squark_k}^*)  
  m_{\quark}^2 m_{\gluino}^2
  \propM_{\Fbar\Fbar\Fbar\Fbar S}
  (\quark,\gluino,\gluino,\quark,\squark_k)
+ \frac{1}{4} P_i^k P_k^m P_m^n P_n^j
   \propS_{SSS} (\squark_k, \squark_m,\squark_n)
\Bigr ]
\phantom{xxx}
\nonumber \\ &&
+ 4 g_3^4 C_q I_q \Bigl [
\delta_i^j   
 \propV_{FFFFS} (\quark,\gluino,\gluino,q_r,\squark_r)
+ \delta_i^j  
  m_{\gluino}^2
  \propV_{F\Fbar\Fbar F S}
  (q,\gluino,\gluino,q_r,\squark_r)
- 2 N_i^j 
  m_{q_r} m_{\gluino}
  \propV_{\Fbar F\Fbar FS}
  (q,\gluino,\gluino,q_r,\squark_r)
\nonumber \\ &&
+ N_i^j N_r^r 
  m_q m_{q_r}
  \propV_{\Fbar FF\Fbar S}
  (q,\gluino,\gluino,q_r,\squark_r)
- 2 \delta_i^j N_r^r 
  m_{q_r} m_{\gluino}
  \propV_{FF\Fbar\Fbar S}
  (q,\gluino,\gluino,q_r,\squark_r)
\nonumber \\ &&  
+ N_i^j N_r^r 
  m_q m_{q_r} m_{\gluino}^2
  \propV_{\Fbar\Fbar\Fbar\Fbar S}
  (q,\gluino,\gluino,q_r,\squark_r)
+ \frac{1}{4} P_i^k P_k^j P_r^s P_s^r 
   \propS_{SSS} (\squark_k, \squark_r,\squark_s)
\Bigr ] .
\label{eq:squarksnogluon}
\eeq
For two-loop diagrams with one gluon propagator,
\beq
\Pi^{(2,1)j}_{i} &=& 
g_3^4 \delta_i^j \bigl \lbrace
C_q C_G [ G_{FF}(\quark, \gluino) 
+ G_{SSFF} (\squark_i,\squark_i,\quark, \gluino) ]
+ [2 C_q - C_G] C_q 
G_{SSFF} (\squark_i,\squark_i,\gluino, \quark)  \bigr \rbrace
\nonumber \\ &&
- g_3^4 
N_i^j
m_{\quark} m_{\gluino}
\bigl \lbrace
C_q C_G [ G_{\Fbar\Fbar}(\quark, \gluino)
+ G_{SS\Fbar\Fbar} (\squark_i,\squark_j,\quark, \gluino) ]
+ [2 C_q - C_G] C_q 
G_{SS\Fbar\Fbar} 
(\squark_i,\squark_j,\gluino, \quark)  \bigr \rbrace
\nonumber \\ &&
+ g_3^4 C_q^2 P_i^k P_k^j 
\left [
G_S(\squark_k) + G_{SSS}(\squark_i,\squark_j,\squark_k)
\right ].
\eeq
Finally, for two-loop diagrams with two or more gluon lines,
\beq
\Pi^{(2,2)j}_i = g_3^4 \delta_i^j 
C_q \Bigl [
C_q F_1 (\squark_i) 
+ C_G F_2 (\squark_i) 
+ C_G F_3(\squark_i, \gluino)
+ I_q \sum_r \lbrace F_3(\squark_i,q_r)
+ F_4(\squark_i,\squark_r) \rbrace  
\Bigr ] .
\label{eq:squarkstwogluon}
\eeq
In eqs.~(\ref{eq:squarksnogluon})-(\ref{eq:squarkstwogluon}), 
indices $r,s$ are used when
all 12 left-handed and right-handed quarks and 12 squark mass eigenstates 
should be summed over. Indices $k,m,n$
are used when only the two
squarks with the same flavor as the external particle
states $\tilde q_i, \tilde q_j$ are summed over. 
The quark corresponding to the external squark flavor is called $q$.

The squark pole squared mass $M^2_{\tilde q_i}$ can now be obtained 
following the discussion of the previous section; 
see eq.~(\ref{eq:poleresultgen}).
This yields: 
\beq
M^2_{\tilde q_i} &=& m^2_{\tilde q_i} 
+ \frac{1}{16 \pi^2} \widetilde \Pi^{(1)i}_{i} 
+ \frac{1}{(16 \pi^2)^2} \Bigl [
\widetilde \Pi^{(2,0)}_{i} 
+ \widetilde \Pi^{(2,1)}_{i} 
+ \widetilde \Pi^{(2,2)}_{i} 
+ \sum_{j\not= i} \widetilde \Pi^{(1)j}_{i} \widetilde \Pi^{(1)i}_{j}/
     (m_{\tilde q_i}^2 - m_{\tilde q_j}^2) \Bigr ],
\label{eq:noname}
\eeq
with no sum on $i$ implied. Within the approximation specified above for
sfermions in the MSSM, $j$ can actually take on only one value for a given
$i$, and the sfermions that mix are always non-degenerate. Here
$\widetilde \Pi^{(1)j}_{i}$
is obtained from $\Pi^{(1)j}_{i}$ by taking
$s \rightarrow m^2_{\tilde q_i}$. Also,
$\widetilde \Pi^{(2,1)}_i$ is obtained from
$\Pi^{(2,1)i}_i$ by replacing the functions
$G_{SSS}$, $G_{SSFF}$ and $G_{SS\Fbar\Fbar}$ by 
$\widetilde G_{SSS}$,
$\widetilde G_{SSFF}$, and
$\widetilde G_{SS\Fbar\Fbar}$, and 
$\widetilde \Pi^{(2,2)}_i$ is obtained from 
$\Pi^{(2,2)i}_i$ by replacing the functions 
$F_{1,2,3,4}$ by $\widetilde F_{1,2,3,4}$, and 
\beq
\widetilde \Pi^{(2,0)}_i =
\lim_{s \rightarrow m_{\tilde q_i}^2} \left [
\Pi^{(2,0)i}_i
+ \widetilde \Pi^{(1) i}_{i} 2 g_3^2 C_q \lbrace
  \propB_{FF}' (q, \tilde g) - N_i^i m_q m_{\tilde g}
  \propB_{\Fbar\Fbar}' (q, \tilde g)\rbrace \right ] .
\label{eq:pitildetwozero}
\eeq

I have checked that this result for the squark pole mass is invariant
under renormalization group evolution of the parameters, up to terms of
three-loop order. That check is somewhat 
messy when non-trivial mixing is involved,
and so will not be presented here in the general case. Instead, it will be
shown explicitly in two simplified special cases in the next section.

\section{Simple examples}\label{sec:examples}
\setcounter{equation}{0}
\setcounter{footnote}{1}

\subsection{Squarks without mixing}
\label{subsec:unmixedsquarks}

In this subsection, I consider a simple (probably non-realistic) example,
to demonstrate the typical
size of the two-loop contribution to the squark pole
masses. As an approximation, suppose that all squarks are degenerate in 
mass, with therefore no mixing,
and that all of the quark masses can be neglected. 
Then, reorganizing the results of the previous section by
common group theory factor rather than number of gluon propagators,
one obtains for the pole squared mass of a given squark $\tilde q_i$:
\beq
M^2_{\tilde q_i} &=& m^2_{\tilde q_i} 
+ \frac{1}{16\pi^2} g_3^2 C_q 
\widetilde \Pi^{(1)} ({\tilde q_i},{\tilde g})
\nonumber \\ &&
+ \frac{1}{(16 \pi^2)^2} g_3^4 C_q \Bigl [
C_q \widetilde \Pi^{(2a)} ({\tilde q_i},{\tilde g})
+ C_G \widetilde \Pi^{(2b)}({\tilde q_i},{\tilde g}) 
+ I_q \sum_r \widetilde \Pi^{(2c)} ({\tilde q_i},{\tilde g},{\tilde q_r}) 
\Bigr ] + \ldots
\label{eq:Mnomixing}
\eeq
where
\beq
\widetilde \Pi^{(1)} (x,y) &=& 
\propA_{S} (x) 
+ 2 \propB_{FF} (0,y) 
+ \propB_{SV}(x,0)
\label{eq:Pionenomixing}
\\
\widetilde \Pi^{(2a)} (x,y) &=&
\propX_{SSS}(x,x,x) 
+ \propS_{SSS}(x,x,x)
+ 4 y \propM_{F\Fbar\Fbar F S}(0,y,y,0,x) 
+ 4 \propV_{FFFFS} (y,0,0,y,x)
\nonumber \\ 
&&
+ 2 \propW_{SSFF}(x,x,0,y) 
+ G_{S}(x) + \widetilde G_{SSS}(x,x,x) + 2 \widetilde G_{SSFF} (x,x,y,0)
+ \widetilde F_1 (x) 
\nonumber \\ 
&&
+ 2 \Pi^{(1)} (x,y) \,\propB'_{FF}(0,y)
\\
\widetilde \Pi^{(2b)} (x,y) &=&
- \frac{1}{2} \propS_{SSS} (x,x,x) 
- 2 y \propM_{F\Fbar\Fbar FS}(0,y,y,0,x) 
+ G_{FF} (0,y) 
+ \widetilde G_{SSFF} (x,x,0,y) 
\nonumber \\ 
&&
- \widetilde G_{SSFF} (x,x,y,0)
+ \widetilde F_2 (x)  
+ \widetilde F_3 (x,y) 
\\
\widetilde \Pi^{(2c)} (x,y,z) &=&
\propS_{SSS}(x,z,z)
+ 4 \propV_{FFFFS} (0,y,y,0,z) +
4 y \propV_{F\Fbar\Fbar FS}(0,y,y,0,z) 
+ \widetilde F_3 (x,0) 
+ \widetilde F_4 (x,z) 
.\phantom{xxx}
\label{eq:Pitwocnomixing}
\eeq
All of the basis integral functions on the right-hand sides of these
equations are to be evaluated at $s=x$.
This example has the virtue that all of these functions can be given
analytically in terms of polylogarithms, with the result (in the 
$\DRbarprime$ scheme):
\beq
\widetilde \Pi^{(1)} (x,y) &=& 
2 x [1 + \ln(y/x) + (1-y/x)^2 \ln(1-x/y)] + 6 y - 4 y \lnbar y
\label{eq:pionenomixing}
\\
\widetilde \Pi^{(2a)} (x,y) &=& 8 (x-y)^2 M(0,0,x,y,0) - 8 (x-y) y M(0,y,y,0,x)
+ (24 x - 8 y - 12 y^2/x) {\rm Li}_2(1-x/y)
\nonumber \\
&& 
+ (1-y/x)^2 (2 x + 4 y - 4 y^2/x) \ln^2(1-x/y)
+ 4 (1-y/x) [6x - 2 y + 4 y^2/x - (x+y) \lnbar x 
\nonumber \\
&& 
+ (x-y - 2 y^2/x) \lnbar y] \ln(1-x/y)
+ (14 x - 4 y) \ln^2(x/y) + 8 [y \ln(x/y) + 3x - y + y^2/x] \lnbar y 
\nonumber \\
&& 
+ 24 (y-x) \lnbar x
+ [24  \zeta(3) - 6  + 16 \pi^2 (1 - \ln2)] x 
- (60 + 8 \pi^2/3) y + (2 \pi^2 - 12) y^2/x
\\
\widetilde \Pi^{(2b)} (x,y) &=&
4 (x - y) [(x + y) M(0, x, y, 0, y) + y M(0, y, y, 0, x)]
-2 (x-y)^2 [2 M(0,0,x,y,0) + M(0,0,y,y,0)]\phantom{xxx}
\nonumber \\
&& 
+ (2x - 12 y + 12 y^2/x) {\rm Li}_2(1-x/y)
+ 8 (y-x) [f(\sqrt{y/x}) + (1-y/x) \ln^2(1-x/y)]
\nonumber \\
&&
+ [ (10x - 8 y) \lnbar x + (44 y - 16 x - 30 y^2/x) \lnbar y
    + 19 x - 80 y + 61 y^2/x] \ln(1-x/y) 
\nonumber \\
&&
+ (2y - x) \lnbar^2 x + (8x - 4 y) \lnbar x \lnbar y 
+ (20 y - 7 x) \lnbar^2 y - (25 x + 16 y) \lnbar x + (19x - 66y) \lnbar y
\nonumber \\
&&
+ x [21 -12 \zeta(3) - 26 \pi^2/3 + 8 \pi^2 \ln 2]
+ y (123 + 14 \pi^2/3 - 2 \pi^2 y/x)
\\ 
\widetilde \Pi^{(2c)} (x,y,z) &=&
[(x - y) (y - z) (x y - 5 y^2  + 3 x z + y z)/x y^2] \lbrace
2 {\rm Li}_2 ([y-z]/[y-x]) + \ln^2 (1-x/y) \rbrace
\nonumber \\
&&
+  [2 (y-z) (5y-z)/x] {\rm Li}_2 (1-y/z)
+ [8 z - 4 z (x+z)/y + 6 x z^2/y^2] {\rm Li}_2 (1-x/z)
\nonumber \\
&&
+ 8 (x+z) f(\sqrt{z/x})
+ 
[6 z +  2 y z/x - 16 y^2/x +
(4 z + 12 y z/x - 2 z^2/x - 6 z^2/y) \ln(z/y)
\nonumber \\
&&
+ (10 y^2/x - 2 y) \lnbar y]
(1-x/y) \ln(1-x/y)
+ \lbrace 
[2 z (2 x/y + 2 z/y - 3 x z/y^2)] \ln(z/x)
\nonumber \\
&&
+ (1-z/x)(10 y - 7 x - 7 z + 6 x z/y) \rbrace \ln(1-x/z)
+ (16 y + 4 z - 6 x z/y) \ln(y/z)
+ (x-6 y) \lnbar^2 y
\nonumber \\
&&
- x \lnbar^2 x
+ [4 z + (y-z)(5y-z)/x + 2 z (x+z)/y - 3 x z^2/y^2] \ln^2(y/z) 
+ (2 x + 26 y + 4 z) \lnbar z
\nonumber \\
&&
+ (9x + 8z) \ln(x/z) 
-7 x - 40 y - 3 z
+ (2 \pi^2/3 ) (2z (x+z)/y -x - 2 z - 3 x z^2/y^2)
.
\label{eq:pitwocnomixing}
\eeq
Ref.~\cite{TSIL} gave analytic formulas for the master integral cases 
$M(0,0,x,y,0)$, $M(0,y,y,0,x)|_{s=x}$. The integral 
$M(0,x,y,0,y)|_{s=x}$ can be reduced using recurrence
relations to results found in \cite{Fleischer:1998dw}, 
\cite{Jegerlehner:2003py}, and was given in the 
present notation in \cite{TSIL}.
Also, $M(0,0,y,y,0)$ was originally found in \cite{Broadhurst:1987ei}
and listed in the present notation in ref.~\cite{evaluation}.
The function $f(z)$ is defined in 
eq.~(\ref{eq:definefunky}) of the present paper.

The renormalization group scale independence of
the pole mass $M_{\tilde q_i}$ can now be checked. 
The requirement $dM_{\tilde q_i}^2/dQ = 0$ amounts to:
\beq
- \beta^{(1)}_{m^2_{\tilde q_i}} &=&
g_3^2 C_q\,
Q \frac{\partial}{\partial Q}\widetilde \Pi^{(1)} ({\tilde q_i},{\tilde g}) 
\label{eq:mione}
\\
- \beta^{(2)}_{m^2_{\tilde q_i}} &=& g_3^4 C_q \,
Q \frac{\partial}{\partial Q} \Bigl [
C_q \widetilde \Pi^{(2a)} ({\tilde q_i},{\tilde g}) 
+ C_G \widetilde \Pi^{(2b)} ({\tilde q_i},{\tilde g}) 
+ I_q \sum_r \widetilde \Pi^{(2c)} ({\tilde q_i},{\tilde g},{\tilde q_r}) 
\Bigr ]
\nonumber \\ 
&&
+ g_3^2 C_q \Bigl [ 
\frac{2}{g_3} \beta^{(1)}_{g_3}
+ 2 m_{\tilde g}
\beta^{(1)}_{m_{\tilde g}}  \frac{\partial}{\partial m^2_{\tilde g}}
+ \beta^{(1)}_{m_{\tilde q_i}^2}  \frac{\partial}{\partial m^2_{\tilde q_i}}
\Bigr ] \widetilde \Pi^{(1)}
({\tilde q_i},{\tilde g})
\label{eq:mitwo}
\eeq
\end{widetext}
(with no sum on $i$),
where
\beq
Q\frac{dX}{dQ} \equiv \beta_X
= \frac{1}{16\pi^2} \beta^{(1)}_X + \frac{1}{(16\pi^2)^2} \beta^{(2)}_X 
+ \ldots ,
\eeq
are the beta functions of running parameters 
$X = g_3, m_{\tilde g}, m_{\tilde q_i}^2$.
Equations (\ref{eq:mione})-(\ref{eq:mitwo}) can be checked using the
results
\beq
\beta^{(1)}_{g_3} &=& g_3^3 (- 3 C_G + 2 N_f I_q),
\label{eq:betagone}
\\
\beta^{(2)}_{g_3} &=& g_3^5 [- 6 C_G^2 + (4 C_G + 8 C_q) N_f I_q],
\\
\beta^{(1)}_{m_{\tilde g}} &=& g_3^2 (- 6 C_G + 4 N_f I_q) m_{\tilde g},
\\
\beta^{(2)}_{m_{\tilde g}} &=& 
g_3^4 [- 24 C_G^2 + (16 C_G + 32 C_q) N_f I_q ] m_{\tilde g},
\phantom{xxxx}
\\
\beta^{(1)}_{m_{\tilde q_i}^2} &=& -8 g_3^2 C_q m_{\tilde g}^2,
\\
\beta^{(2)}_{m_{\tilde q_i}^2} &=& g_3^4 C_q \Bigl [ 
(-80 C_G + 48 C_q + 48 N_f I_q) m_{\tilde g}^2
\nonumber \\ &&
+ 8 I_q \sum_r m_{\tilde q_r}^2 \Bigr ]
,
\label{eq:betamsquarktwo}
\eeq
with $2 N_f$ the number of quark/squark chiral
superfields  (12 in the MSSM), and
\beq
Q \frac{\partial}{\partial Q}\widetilde \Pi^{(1)} (x,y) &=& 8y ,
\\
\frac{\partial}{\partial x}\widetilde \Pi^{(1)} (x,y) &=& 
2 - 2y/x + 2\ln(y/x) 
\nonumber \\ && 
+ 2 (1 - y^2/x^2) \ln (1-x/y)
,
\\
\frac{\partial}{\partial y}\widetilde \Pi^{(1)} (x,y) &=&
4 + 4 (y/x-1) \ln (1-x/y) 
\nonumber \\ &&
- 4 \lnbar y, 
\\
Q \frac{\partial}{\partial Q}\widetilde \Pi^{(2a)} (x,y) &=&
-32 y - 16 y^2/x + 16 y \ln(y/x) 
\nonumber \\ &&
\!\!\!\!\!\!\!\! 
+ 16 y (1-y^2/x^2) \ln(1-x/y)
,
\\
Q \frac{\partial}{\partial Q}\widetilde \Pi^{(2b)} (x,y) &=&
12 x + 164 y + 12 x \ln(y/x) 
\nonumber \\ &&
\!\!\!\!\!\!\!\! 
+ (12 x - 72 y + 60 y^2/x) \ln(1-x/y) 
\nonumber \\ &&
\!\!\!\!\!\!\!\! 
- 72 y \lnbar y
,
\phantom{xxx}
\\
Q \frac{\partial}{\partial Q}\widetilde \Pi^{(2c)} (x,y,z) &=&
-4 x - 52 y  - 8 z + 4 x \ln(x/y) 
\nonumber \\ &&
\!\!\!\!\!\!\!\!  
+ (24 y - 4 x - 20 y^2/x) \ln(1-x/y)\phantom{xx}
\nonumber \\ &&
\!\!\!\!\!\!\!\!  
+ 24 y \lnbar y, 
\eeq
which in turn follow directly from 
eqs.~(\ref{eq:pionenomixing})-(\ref{eq:pitwocnomixing}), noting
that the master integral basis function $M(x,y,z,u,v)$ has no
explicit dependence on $Q$.

\begin{figure}[t]
\centering
\includegraphics[width=8.6cm]{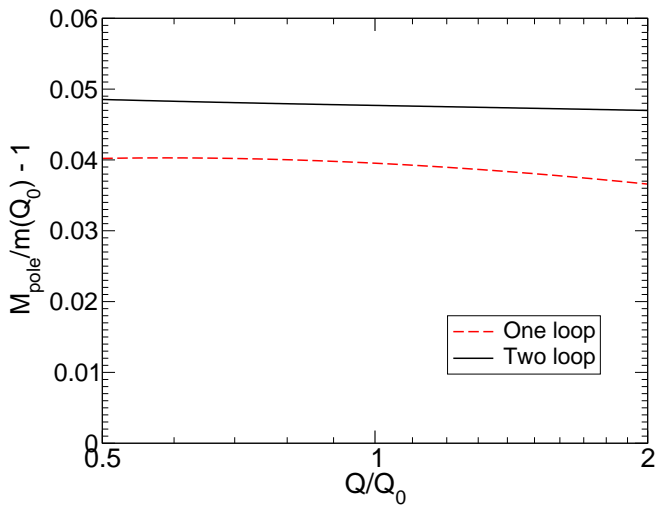}
\caption{\label{fig:scaledep} 
Dependence of the calculated pole mass for degenerate
squarks on the choice of renormalization
scale. 
The input parameters are defined by running gluino and
squark masses taken to be equal at an input renormalization
scale $Q_0 = m_{\tilde q}(Q_0) = m_{\tilde g}(Q_0)$, 
with $\alpha_S(Q_0) = 0.095$.
These parameters are then run to new scales $Q$, where the pole mass is
recomputed. The vertical axis is the fractional change in the squark pole
mass compared to the squark running mass at the input scale.
Six squark families are assumed as in the MSSM, but quark masses, 
squark mixing, and non-SUSYQCD effects are neglected in this graph.}
\end{figure}

This scale independence thus holds up to terms of three-loop order. To
illustrate it in practice, consider the even more special case that
the gluino and all squarks have equal running masses at an 
input renormalization scale $Q_0$ given by the same value, so that
$Q_0 = m_{\tilde g}(Q_0) = m_{\tilde q}(Q_0)$. 
Figure \ref{fig:scaledep} then shows the scale dependence of the 
squark pole
mass as calculated from eqs.~(\ref{eq:Mnomixing}) and 
(\ref{eq:pionenomixing})-(\ref{eq:pitwocnomixing}).
To make this graph, the running parameters
$g_3$, $m_{\tilde g}$, and $m_{\tilde q}$ are each run from the input
scale $Q_0$ to a new scale $Q$, using their two-loop renormalization
group equations (\ref{eq:betagone})-(\ref{eq:betamsquarktwo}). 
Here I have put in the MSSM values,
namely $C_G = 3$, $C_q = 4/3$, $I_q = 1/2$, and $N_f=6$, and taken 
$\alpha_S(Q_0) = g_3^2(Q_0)/4\pi = 0.095$. 
At the scale $Q$, the
pole mass is recomputed, and the 
quantity $M_{\tilde q}/m_{\tilde q}(Q_0) - 1$
is shown; in the ideal case of an exact calculation
the resulting line would be exactly horizontal. 
The two-loop result has 
a slightly improved scale dependence, as expected, but the
difference between the two-loop result and the one-loop result is actually
much larger than the scale dependence of the latter. This demonstrates
that the scale dependence does not give an adequate estimate of the
theoretical error of the calculation.
\begin{figure}[t]
\centering
\includegraphics[width=8.6cm]{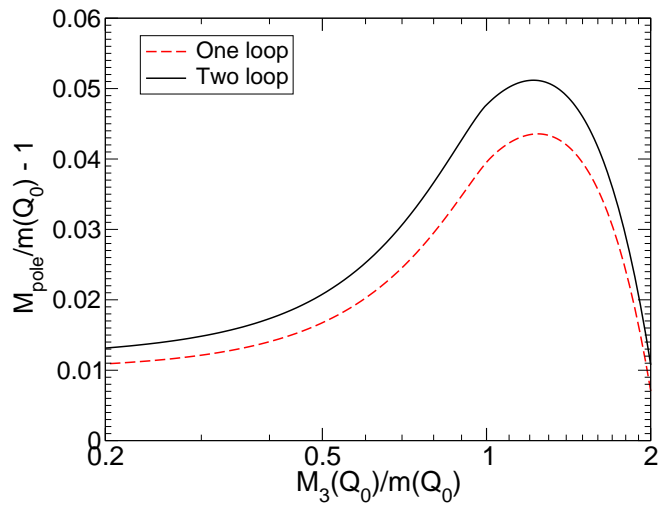}
\caption{\label{fig:gluinovar} 
The SUSYQCD
corrections to the squark pole mass, as in figure \ref{fig:scaledep},
but now as a function of the varying ratio 
of the tree-level running gluino mass 
$M_3(Q_0) \equiv m_{\tilde g}(Q_0)$ to the running 
squark masses $Q_0 = m_{\tilde q}(Q_0)$. 
The squark pole mass is computed at the renormalization scale $Q_0$.}
\end{figure}

The result at $Q=Q_0$ is:
\beq
M^2_{\tilde q} &=& 
m^2_{\tilde q} \Bigl [
1
+ \frac{\alpha_S}{4 \pi} \left (\frac{32}{3} \right )
+ \left (\frac{\alpha_S}{4 \pi}\right )^2 \Bigl (
\frac{112}{3}
\nonumber \\
&&
+ \frac{664 \pi^2}{27}
+ \frac{32 \pi^2 \ln 2}{9}  
- \frac{16 \zeta(3)}{3}
\Bigr ) \Bigr ]
\\
&=&
m^2_{\tilde q} \left [ 1 + 0.849 \alpha_S + 1.89 \alpha_S^2 \right ] .
\eeq
There are no logarithms
here, since there is only one mass scale,
so the result gives some 
idea of the typical intrinsic size of the two-loop corrections.
The one-loop correction is an increase of order 4\% in the pole mass
compared to the running mass evaluated at itself, while the two-loop
correction adds an additional amount of order 1\%. 

More generally, fig.~\ref{fig:gluinovar}
shows the one-loop and two-loop corrections to the squark masses,
calculated as above,
but now varying the running gluino mass $M_3(Q_0) \equiv m_{\tilde g}(Q_0)$
at the fixed renormalization scale $Q_0 = m_{\tilde q}(Q_0)$. 
The two-loop part of the correction
is seen to be largest when the
squark masses are slightly less than the gluino mass, and does not 
exceed 1\% over the indicated range. 

In nearly all realistic models
of supersymmetry breaking, $M_3/m_{\tilde q} < 1.5 $ at the TeV scale
for the squarks of the first two families.
For $M_3(Q_0)$ much larger than 
$2 m_{\tilde q}(Q_0)$, there are large negative loop corrections to the
squark pole mass from gluino loops. This can happen for top and bottom
squarks in the MSSM, in which case the top and bottom Yukawa couplings and
scalar cubic couplings must be included to give a reliable result.

\subsection{Squarks in the supersymmetric limit}

Let us next consider the supersymmetric limit for squarks for an $SU(n)$
supersymmetric Yang-Mills theory with gauge coupling $g$. The gaugino
mass vanishes, and $N_f$ flavors of quarks and squarks
obtain their masses solely from a superpotential
\beq
W = \sum_{i=1}^{N_f} m_{i} \overline Q^i Q_i .
\eeq
Here, $Q_i$ are chiral superfields 
transforming in the fundamental representation, and 
$\overline Q_i$ in the anti-fundamental representation of the gauge group. 
Specializing the results of section IV to this case, I 
obtain two-loop squark pole squared masses:
\beq
M_i^2 = m_i^2 
+ \frac{1}{16\pi^2}\widetilde \Pi^{(1)}_i
+ \frac{1}{(16\pi^2)^2}\widetilde \Pi^{(2)}_i
\label{eq:susylimit}
\eeq
where
\beq
\widetilde \Pi^{(1)}_i &=& g^2 C_q m_i^2 (8 - 4 \lnbar m_i^2)
\\
\widetilde \Pi^{(2)}_i &=& g^4 C_q \Bigl (
C_q m_i^2 
\Bigl [40 \pi^2/3 - 16 \pi^2 \ln 2 
\nonumber \\ &&
+ 24 \zeta(3) 
-28 
- 8 \lnbar m_i^2 + 8 \lnbar^2 m_i^2 \Bigr ] 
\nonumber \\ &&
+ C_G m_i^2 \Bigl [ 66 - 4 \pi^2 
+ 8 \pi^2 \ln 2 
- 12 \zeta(3) 
\nonumber \\ &&
- 36 \lnbar m_i^2
+ 6 \lnbar^2 m_i^2 \Bigr ]
\nonumber \\ &&
+ 2 I_q \sum_{j=1}^{N_f} h(m_i^2,m_j^2)
\Bigr ),\phantom{xxxxx}
\label{eq:susylimittwo}
\eeq
in which
\beq
h(x,y) &=& 4 (x+y) [{\rm Li}_2(1-x/y) - \pi^2/6] 
\nonumber \\ &&
+ x \bigl [ 2 \lnbar^2y 
         - 4 \lnbar x \lnbar y + 12 \lnbar x 
\nonumber \\ &&
             + 16 f (\sqrt{y/x}) - 22 \bigr ],
\\
h(x,0) &=& x (-22 - 4 \pi^2/3 + 12 \lnbar x - 2 \lnbar^2 x),\phantom{xxx}
\\
h(0,y) &=& 0,
\eeq
with $f(z)$ defined by eq.~(\ref{eq:definefunky}).

The renormalization group invariance of this pole mass result
now follows from
\beq
\beta^{(1)}_{g} &=& g^3 (- 3 C_G + 2 N_f I_q),
\\
\beta^{(1)}_{m_i} &=& -4 g^2 C_q m_{i},
\\
\beta^{(2)}_{m_i} &=& g^4 C_q (8 C_q - 12 C_G + 8 N_f I_q) m_{i}.
\eeq

Since supersymmetry is unbroken in this example, the result of
eqs.~(\ref{eq:susylimit})-(\ref{eq:susylimittwo}) must be equally valid
for the quark pole squared masses as for the squark pole squared masses
derived directly here.

\subsection{Gauge mediation of supersymmetry breaking}

As another example application, I show how to reproduce 
the result of gauge mediation of
supersymmetry breaking to MSSM scalars by specializing the results above. 
In this case, the two-loop order result gives the
leading effect, found in
ref.~\cite{Dimopoulos:1996gy}. 
(A derivation in terms of individual diagrams in Feynman gauge, 
perhaps useful for comparison with the treatment here, was later given 
in \cite{Martin:1996zb}.)
Suppose that there exists a new, vector-like, heavy ``messenger"
quark/squark pair (not part of the MSSM). The heavy quark is taken to have
a Dirac mass $m_Q$, and its scalar superpartners have a squared-mass 
matrix of the form:
\beq
\begin{pmatrix}
m_Q^2 & \Delta\\ 
\Delta &m_Q^2
\end{pmatrix},
\eeq
where $\Delta$ is a supersymmetry-breaking effect. Diagonalizing this
mass matrix according to eq.~(\ref{eq:diagsquarks}) leads to eigenvalues
$m_{\tilde Q_\pm}^2 = m_Q^2 \pm \Delta$, with a mixing angle of $\pi/4$.
This induces masses for
the ordinary squarks of the MSSM, which can be treated as 
massless in leading order. From 
eqs.~(\ref{eq:squarksnogluon})-(\ref{eq:pitildetwozero}), 
the result for the ordinary squark masses is:
\beq
M^2_{\tilde q_i} &=& \frac{g_3^4 C_q I_q}{(16 \pi^2)^2} 
\Bigl [
4 \propV_{FFFFS}(0,0,0,Q,\widetilde Q_+)
\nonumber \\
&&
\!\!\!\!\!\!\!\!\!\!
+ 4 \propV_{FFFFS}(0,0,0,Q,\widetilde Q_-)
+ 2 \propS_{SSS} (0, \widetilde Q_+, \widetilde Q_-)
\nonumber \\
&&
\!\!\!\!\!\!\!\!\!\! 
+ 2 \widetilde F_3 (0,Q)
+ \widetilde F_4 (0, \widetilde Q_+)
+ \widetilde F_4 (0, \widetilde Q_-) \Bigr ] .
\eeq
Then, one can use eqs.~(\ref{eq:Fthreeoy})-(\ref{eq:Ffouroy}) and 
the results valid for $s=0$:
\beq
&&\propV_{FFFFS}(0,0,0,x,y) = 
\frac{1}{x-y} \Bigl [
-2 y I(0,x,y) 
\phantom{xxx}
\nonumber \\ 
&&
\qquad\qquad
- 2 x y \lnbar x \lnbar y + 2 y (x+y) \lnbar y
\phantom{xxxxxx}
\nonumber \\ 
&&
\qquad\qquad
+ 2 x (x+y) \lnbar x 
- 3 x^2 - 3 y^2 - 4 x y \Bigr ],
\phantom{xxxx}
\\
&&\propS_{SSS}(0,x,y) = -I(0,x,y),
\eeq
where 
\beq
I(0,x,y) &=& (x-y) [{\rm Li}_2(1-x/y) + (\lnbar y)^2/2] 
\nonumber \\ 
&&
\!\!\!\!\!\!\!\!\!\!\!\!\!\!\!\!\!\!\!\!
- x \lnbar x \lnbar y + 2 x\lnbar x + 2 y \lnbar y - 5 (x+y)/2
.
\phantom{xxxxx}
\eeq
The result for $M_{\tilde q_i}^2$ is
equivalent to the one originally given in
ref.~\cite{Dimopoulos:1996gy}. It is not hard to generalize this to the
$SU(2)_L$ and $U(1)_Y$ gauge groups to obtain
the full set of predictions for MSSM squark and slepton
masses in gauge-mediated supersymmetry breaking.

\section{Outlook}
\setcounter{equation}{0}
\setcounter{footnote}{1}

In this paper, I have found the two-loop contributions to scalar boson
self-energies, and thus pole masses, in a general gauge theory with
massless (or light) gauge bosons. These results should apply directly to
heavy scalars in perturbative models of physics beyond the Standard Model,
provided the $W$ and $Z$ masses can be neglected compared to the dominant
mass scales in the problem. This is quite likely to be a good
approximation, for example, for the squarks and sleptons in supersymmetric
theories, where the difference between the full two-loop result and the
one reported here is suppressed by $m_Z^2/m_{\rm sfermion}^2$ multiplied
by an expansion coefficient that is typically a fraction of unity, as well
as a weak interaction two-loop factor. 

In section \ref{sec:squarks}, I have given the SUSYQCD contributions for
squark masses. However, it should be emphasized that the computations of
all two-loop contributions to all of the the sfermion pole masses in this
approximation have been reduced to an exercise (admittedly tedious, but
certainly amenable to automation by a symbolic manipulation program) in
substitution of running couplings constants and tree-level masses into the
formulas here and in ref.~\cite{Martin:2003it}, followed by numerical
computation of basis integrals using a program such as \cite{TSIL}. In
doing so, the Higgs scalar and the electroweak vector boson sectors can be
treated in an approximation where the effects of electroweak symmetry
breaking are consistently neglected in the two-loop parts. The one-loop
part can of course be treated exactly using the formulas in
\cite{Pierce:1996zz}. 

A convincing guess as to the likely size of the remaining theoretical
errors is difficult to obtain. The results of the example in subsection
\ref{subsec:unmixedsquarks} may suggest that three-loop effects on squark
masses are usually less than a few tenths of a percent, but the limited
available data on the convergence of the perturbative expansion here is
not clearly in support of this conjecture. Also, the scale dependence of
the result, although quite mild, has often been seen to underestimate the
theoretical error. It should be noted that there will also be substantial
sources of irreducible experimental error, notably uncertainties in
$\alpha_S$, the gluino mass, and the other superpartner masses. It seems
likely that global fits to many different observables will be necessary in
order to extract the parameters of the underlying Lagrangian. Clearly,
there will be many challenges to overcome to go from future experimental
data to a clear and precise understanding of the origin of supersymmetry
breaking. 

\section*{Appendix}
\label{appendixA} \renewcommand{\theequation}{A.\arabic{equation}}
\setcounter{equation}{0}
\setcounter{footnote}{1}

In this Appendix, I note the existence of some identities for
two-loop self-energy integral basis function 
that are useful for
deriving some of the formulas of section \ref{sec:selfenergy}.
These include equations (A.14)-(A.20) 
of ref.~\cite{Martin:2003it}, and
\beq
0 &=& (s-x) [T(x,0,0) - U(x,0,0,0) ] 
+ s B(0,x)^2 
\phantom{xx}
\nonumber \\ &&
+ [s - 3 x + (1 + s/x) A(x)] B(0,x) 
\nonumber \\ &&
- x +2 s - 2 A(x) 
+ A(x)^2/x .
\eeq
Also needed are the values \cite{Broadhurst:1987ei,Gray:1990yh}
in the threshold limit $s \rightarrow x$:
\beq
B(0,x) &=& 2 - \lnbar x 
\\
T(x,0,0) &=& \pi^2/3 - 1/2 - \lnbar x + (\lnbar x)^2/2
\phantom{xxxx}
\\
U(x,0,x,x) &=& \frac{11}{2} - \frac{2 \pi^2}{3} - 3\lnbar x + 
\frac{1}{2}\lnbar^2 x 
\\
U(x,0,0,0) &=& \frac{11}{2} + \frac{\pi^2}{3} - 3\lnbar x + 
\frac{1}{2}\lnbar^2 x 
\\
M(0,x,x,0,x) &=& [\pi^2 \ln 2 - 3 \zeta(3)/2]/x ,
\eeq
and the pseudo-threshold expansions \cite{Berends:1997vk}:
\beq
S(x,y,y) &=& S_0(x,y,y) + (1-s/x) S_1(x,y,y) 
\nonumber \\
&&
+ (1-s/x)^2 S_2 (x,y,y) + 
\ldots,
\eeq
where
\beq
S_0 (x,y,y) &=& 
-[(x-y)^2/x] [{\rm Li}_2(1-y/x) + \pi^2/6]
\nonumber \\ &&
+ x (3 \lnbar x - \lnbar^2 x -3/4)/2 
\nonumber \\ &&
+ y [\lnbar^2 x -4 + 5 \lnbar y 
- \lnbar x  
- 2 \lnbar x \lnbar y 
\nonumber \\ &&
- (y/2x) \ln^2 (y/x)]
\\
S_1 (x,y,y) &=&
y (1-y/x) [{\rm Li}_2 (1-y/x) + \pi^2/6 
\nonumber \\ &&
+ \ln^2(y/x)/2]
-5x/8 + (x/2) \lnbar x 
\nonumber \\ &&
+ y  + y \ln(y/x)
\\
S_2 (x,y,y) &=& S_1(x,y,y) + \frac{x}{2} 
\Bigl [\frac{7}{4} - f(\sqrt{y/x}) -\lnbar x \Bigr ],
\phantom{xxxxx}
\eeq
with $f(z)$ defined in eq.~(\ref{eq:definefunky}),
and
\beq
S_0 (x,x,x) &=& x \left [-\frac{35}{8} + \frac{11}{2} \lnbar x 
- \frac{3}{2} \lnbar^2 x \right ] ,\phantom{xxxx}
\\
S_1 (x,x,x) &=& x \left [\frac{3}{8} + \frac{1}{2} \lnbar x \right ],
\\
S_2 (x,x,x) &=& x \left [ 5/4 - \pi^2/8 \right ],
\eeq
and
\beq
T(x,y,y) &=& -\frac{\partial}{\partial x} S(x,y,y),
\\
T(y,y,x) &=& -\frac{1}{2} \frac{\partial}{\partial y} S(x,y,y),
\\
T(x,x,x) &=& -\frac{1}{3} \frac{\partial}{\partial x} S(x,x,x) .
\eeq

I am grateful to David G. Robertson for valuable conversations and
collaboration on the two-loop self-energy integral computer program TSIL
(ref.~\cite{TSIL}). This work was supported by the National Science
Foundation under Grant No.~PHY-0140129.

\end{document}